\documentclass[%
 aip,
 amsmath,amssymb,
 reprint,%
author-numerical,%
]{revtex4-1}

\usepackage{graphicx}
\usepackage{dcolumn}
\usepackage{bm}
\usepackage{amssymb} 
\usepackage{amsmath} 
\usepackage{color}
\usepackage{float}
\usepackage{natbib}
\setcitestyle{numbers}
\setcitestyle{square}

\usepackage[utf8]{inputenc}
\usepackage[T1]{fontenc}
\usepackage{mathptmx}
\usepackage{etoolbox}

\usepackage{endnotes}

\makeatletter
\def\@email#1#2{%
 \endgroup
 \patchcmd{\titleblock@produce}
  {\frontmatter@RRAPformat}
  {\frontmatter@RRAPformat{\produce@RRAP{*#1\href{mailto:#2}{#2}}}\frontmatter@RRAPformat}
  {}{}
}%
\makeatother
\begin{document}

\preprint{AIP/123-QED}

\title[Computation of generalised magnetic coordinates asymptotically close to the separatrix]{Computation of generalised magnetic coordinates asymptotically close to the separatrix}

\author{S. Benjamin}
\affiliation{ 
Plasma Science and Fusion Center, Massachusetts Institute of Technology, USA
}%
\email{sbenjamin@psfc.mit.edu, ncl2128@columbia.edu}
 
\author{N.C. Logan}%

\author{C. Hansen}
\affiliation{%
Columbia University, USA
}%

\date{\today}

\begin{abstract}
Integrals to calculate generalised magnetic coordinates from an input magnetic flux function asymptotically close to the separatrix are presented, and implemented in the GPEC/DCON code suite. These integrals allow characterisation of the magnetic equilibrium of a diverted tokamak, in magnetic coordinates,  arbitrarily close to the last closed flux surface, avoiding the numerical issues associated with calculating diverging field line integrals near a magnetic x-point. These methods provide an important first step in the development of robust asymptotic equilibrium behaviour for spectral 3D MHD codes at the separatrix.
\end{abstract}

\maketitle

\section{Introduction}
Magnetic coordinates have provided myriad advantages in magnetic confinement fusion plasma research. These coordinates, which adhere to the magnetic geometry of the device, can vastly simplify both the analytical expression of stability problems and their computation. Their application in diverted tokamaks, however, is hindered by the fact that the poloidal magnetic field approaches zero at the magnetic x-point on the last closed flux surface (LCFS). This null point in the poloidal field produces an unavoidable divergence in one of two periodic magnetic angles in tokamak magnetic geometry. The result is that the computational domain of spectral MHD codes must be truncated somewhere inside of the LCFS, and unfortunately global stability calculations retain some degree of sensitivity to the exact point of truncation.\\

In this publication we will discuss this problem in some detail, before presenting a first step to its final solution: reliable and simple analytic integrals that extend the computational domain arbitrarily close to the magnetic separatrix. These integrals have been implemented in the GPEC/DCON code suite \cite{glasser_direct_2016, park_computation_2007, park_importance_2009}. Note the complete resolution of the spectral edge truncation problem requires the development of appropriate asymptotic numerical methods at the separatrix, that will  depend on the particular stability problem being solved.

\section{Background}

The magnetic coordinate representation we will apply is the same as that used in DCON \cite{glasser_direct_2016}. We repeat its formulation here for convenience. Let us assume the existence of a magnetic flux coordinate $\psi$ (as is appropriate in axisymmetry \cite{freidberg_ideal_2014}) such that $\mathbf{B}\cdot \nabla \psi =0$. Due to the divergence-free property of the magnetic field, we can define two magnetic angle coordinates, $\zeta$ and $\theta$ that are periodic around the toroidal and poloidal directions of the tokamak respectively, such that field lines are straight lines in the $\zeta,\theta$ plane, \begin{equation}\label{field_coords}
\mathbf{B} = (\nabla \zeta - q\nabla \theta)\times \nabla \psi, \end{equation}
with the gradient of the field lines given by the q-profile, $$q(\psi) \equiv \frac{\mathbf{B} \cdot \nabla \zeta}{\mathbf{B} \cdot \nabla \theta}.$$ 
The standard stability problem through which we will discuss edge truncation is that of finding the solution for minimum-energy ideal linear perturbations in axisymmetric geometry. To refresh the reader's memory on this problem, consider the energy of a general ideal internal plasma perturbation about an equilibrium state \cite{freidberg_ideal_2014} $\mathbf{J}\times \mathbf{B}=\nabla p$ : 
\begin{figure*}[t]
\includegraphics[width=0.8\textwidth]{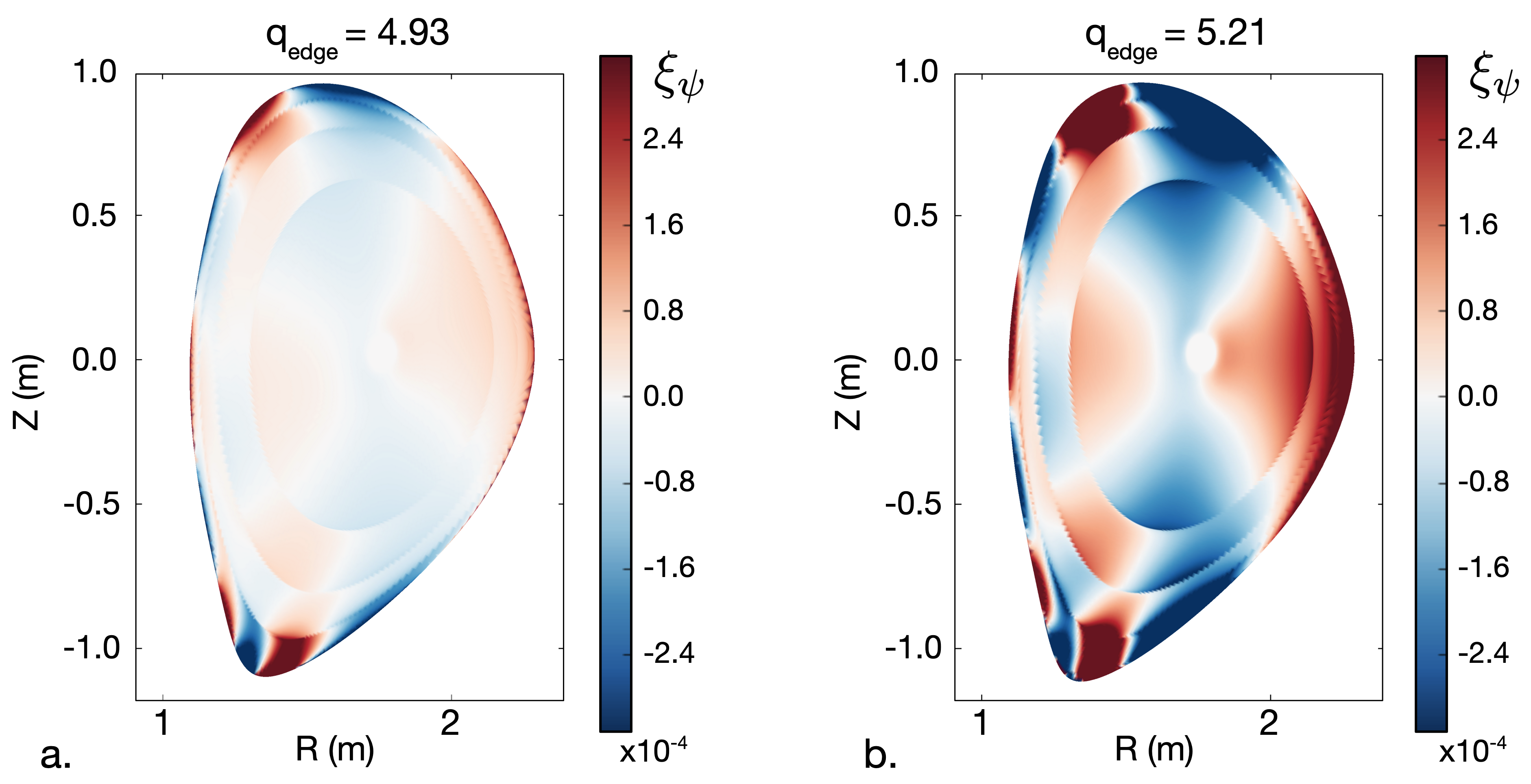}
\caption{\label{fig:Externalkink} DCON calculation of the $n=1$, minimum-$\delta W$ plasma displacement in a single-null DIII-D equilibrium for two different edge truncations. Here $\psi_{edge}= 0.987$ and $0.992$ for panels a and b respectively (3 s.f.). The $q_{edge}=4.93$ displacement is highly localised at the plasma edge, forming an external kink, while the $q_{edge}=5.21$ displacement has greater internal structure.}
\end{figure*}
\begin{align}\label{equ:deltaW_general}
    \delta W = \frac{1}{2\mu_0}\int_V d\mathbf{R} [&\mathbf{B_1}^2+\mathbf{J}\cdot \boldsymbol{\xi}\times\mathbf{B_1}+\nonumber \\ &\mu_0(\boldsymbol{\xi}\cdot\nabla p)(\nabla\cdot\boldsymbol{\xi})+\mu_0\gamma p(\nabla\cdot\boldsymbol{\xi})^2]
\end{align}
where $\mathbf{B_1}=\nabla \times (\boldsymbol{\xi}\times \mathbf{B})$ is the perturbed magnetic field, $\boldsymbol{\xi} = \frac{\partial \mathbf{R}}{\partial \psi}\xi_\psi +\frac{\partial \mathbf{R}}{\partial \zeta}\xi_\zeta +\frac{\partial \mathbf{R}}{\partial \theta}\xi_\theta$ is a small, general plasma displacement in contravariant form, $\mathbf{J}, \mathbf{B}$ and $p$ are the background current, magnetic field, and pressure respectively, and $\gamma$ is thermodynamic ratio of specific heats. Following \cite{glasser_direct_2016}, we can write the contravariant coefficients of $\boldsymbol{\xi}$ in terms of a Fourier basis in magnetic angles: 
$$\xi_j(\psi,\zeta,\theta) =  \sum_{m=-\infty}^\infty \sum_{n=-\infty}^\infty \bar{\xi}_j|_{m,n}(\psi)e^{2\pi i(m\theta-n\zeta)},\hspace{2mm} j=(\psi,\zeta,\theta)$$ where $m$ and $n$ are poloidal and toroidal mode numbers respectively. After substituting the Fourier representation of $\boldsymbol{\xi}$ into Eq. \ref{equ:deltaW_general}, and minimising $\delta W$ by dropping strictly positive terms (eliminating $\xi_\theta$ from the integral \& rendering the perturbations incompressible \cite{freidberg_ideal_2014}), one can rewrite $\xi_\zeta$ in terms of $\xi_\psi$ to arrive at, \begin{align}\label{equ:deltaW_general2}
    \delta W &= \frac{1}{2\mu_0}\int d\psi 
 [\boldsymbol{\Xi}'^\dag \mathbf{F} \boldsymbol{\Xi} + \boldsymbol{\Xi}'^\dag\mathbf{K}\boldsymbol{\Xi} + \boldsymbol{\Xi}^\dag\mathbf{K}^\dag\boldsymbol{\Xi}' + \boldsymbol{\Xi}^\dag\mathbf{G}\boldsymbol{\Xi}], \nonumber 
 \end{align}
 where $\boldsymbol{\Xi}$ is a vector of Fourier amplitudes of $\xi_\psi$ only:
 \begin{align}
 \boldsymbol{\Xi} &= \{\bar{\xi}_\psi|_{m,n}(\psi)|m_{\rm low} \leq m \leq m_{\rm high}\},
\end{align}
and $\mathbf{F},\mathbf{G}$ and $\mathbf{K}$ are matrices of Fourier components of equilibrium quantities \cite{glasser_direct_2016}. We apply the Euler Lagrange equations to the above integral to get a second order matrix ODE:
\begin{align}\label{equ:OilyLagrange}
    \mathbf{L}\boldsymbol{\Xi}(\psi)\equiv
    -(\mathbf{F}\boldsymbol{\Xi}'+\mathbf{K}\boldsymbol{\Xi})'+(\mathbf{K}^\dag\boldsymbol{\Xi}'+\mathbf{G}\boldsymbol{\Xi}) = 0.
\end{align}

Eq. \ref{equ:OilyLagrange} describes the minimum-energy, incompressible perturbations in an axisymmetric magnetic field. It demonstrates how magnetic coordinates have reduced what could have been a very complex 2 or 3D problem (in cylindrical or Cartesian coordinates, for example) to a 1D integration that can be solved using pre-existing adaptive integrators such as LSODE \cite{hindmarsh_odepack_1983}. Applications of Eq. \ref{equ:OilyLagrange} include efficiently testing internal mode ideal stability via the sign of a 1D scalar \cite{glasser_direct_2016}, calculating the total plasma + vacuum $\delta W$ \cite{glasser_direct_2016,glasser_riccati_2018}, calculating the least stable mode structure for resistive wall mode control \cite{bialek_modeling_2001},
calculating plasma response to resonant magnetic perturbation and/or error field correction coils \cite{park_computation_2007, park_importance_2009}, calculating the corresponding neoclassical transport and/or kinetic-MHD stability \cite{logan_neoclassical_2013,park_self-consistent_2017}, and calculating the ideal-region drive to tearing by examining ratios of divergent \& non-divergent components of ideal solutions at rational $q$ surfaces \cite{pletzer_linear_1994,glasser_computation_2016,glasser_robust_2018} where $q(\psi)=m/n$. 

To fully understand the difficulty of edge truncation, the reader must understand that Eq. \ref{equ:OilyLagrange} becomes singular at rational surfaces, due to the incompressibility condition \cite{freidberg_ideal_2014}. To illustrate this, $\mathbf{F}, \mathbf{G}$ and $\mathbf{K}$ can be written in terms of a diagonal matrix $\mathbf{Q}$: 
$$\mathbf{F}=\mathbf{Q\bar{F}Q}\text{, }\text{ }\mathbf{K}=\mathbf{Q\bar{K}}\text{, }\text{ }\mathbf{G}=\mathbf{\bar{G}}\text{, }\text{ }\text{ }\text{ }\text{ } \mathbf{Q}_{m,m'} \equiv (m-nq(\psi))\hspace{0.3mm}\delta_{m,m'},$$
such that when Eq. \ref{equ:OilyLagrange} is reformulated as two coupled first order ODEs \cite{glasser_direct_2016}: 
\begin{align} \label{first_order_EL}
    \mathbf{u}' &= \mathbf{H}\mathbf{u} \text{, where }\mathbf{u} = \begin{pmatrix} \boldsymbol{\Xi} \\
    \mathbf{F}\boldsymbol{\Xi}'+\mathbf{K}\boldsymbol{\Xi}\end{pmatrix},\\
\mathbf{H} &= \begin{pmatrix}
-\mathbf{F^{-1}}\mathbf{K} & \mathbf{F^{-1}} \\
\mathbf{G}-\mathbf{K}^\dag\mathbf{F}^{-1}\mathbf{K} & \mathbf{K}^\dag\mathbf{F}^{-1}\end{pmatrix}
\nonumber\\&=\begin{pmatrix}
-\mathbf{Q^{-1}}\mathbf{\bar{F}^{-1}}\mathbf{\bar{K}} & \mathbf{Q^{-1}}\mathbf{\bar{F}^{-1}}\mathbf{Q^{-1}} \\
\mathbf{\bar{G}}-\mathbf{\bar{K}}^\dag\mathbf{\bar{F}}^{-1}\mathbf{\bar{K}} & \mathbf{\bar{K}}^\dag\mathbf{\bar{F}}^{-1}\mathbf{Q^{-1}}\end{pmatrix},\nonumber
\end{align}
where 
\begin{figure*}[t]
\centering
\includegraphics[width=0.8\textwidth]{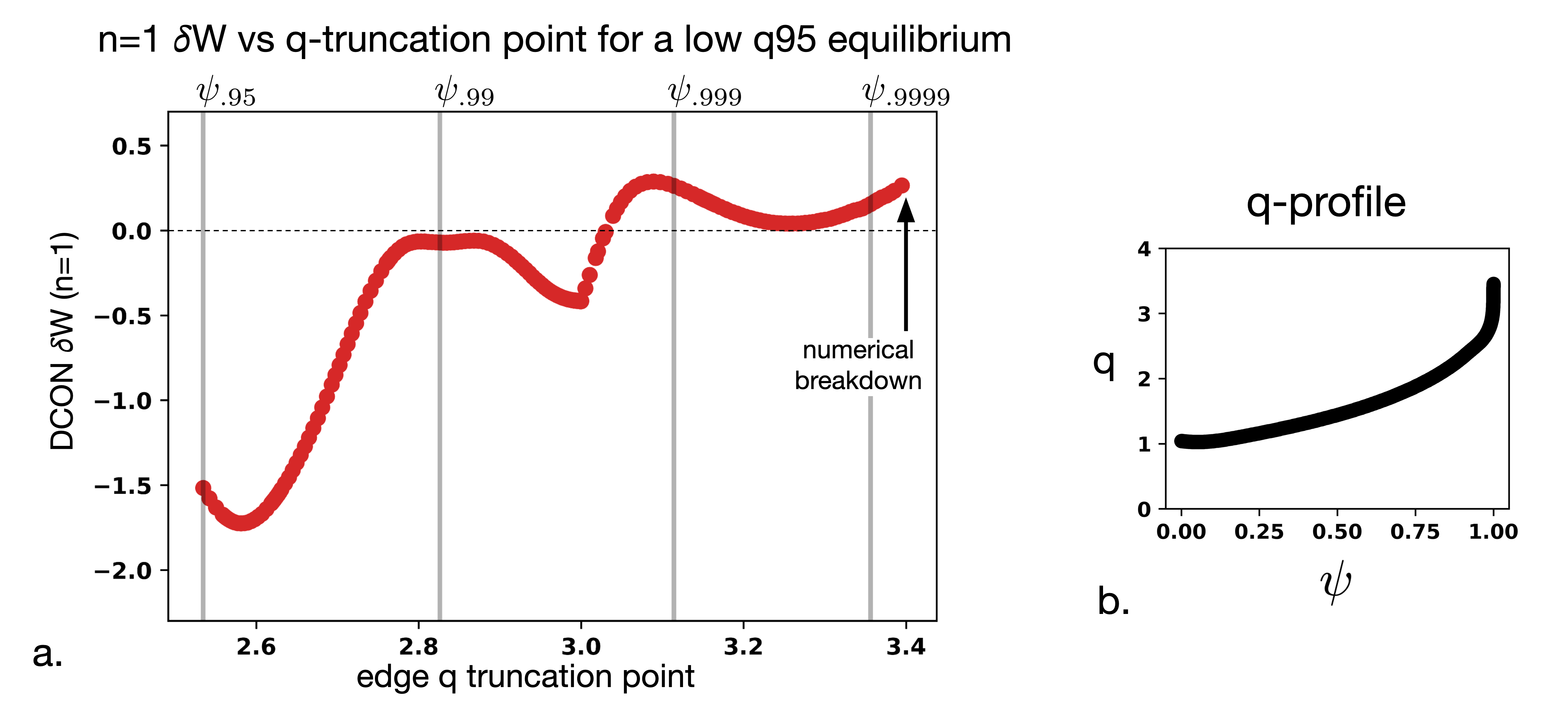}
\caption{\label{DCON_DW_calc} DCON calculation of $\delta W_{n=1}$ for a low q95 H-mode pilot-plant scenario generated in TokaMaker \cite{hansen_tokamaker_2024}. $\psi$ is normalised poloidal flux.}
\end{figure*}
$$\mathbf{Q}^{-1} = \begin{pmatrix}
\frac{1}{m_{\rm low}-nq(\psi)} &   0 &  \hdots  &0 \\
   0 &\frac{1}{(m_{\rm low}+1)-nq(\psi)} &    &0\\
    \vdots&   & \ddots    &     \vdots \\
    0&   0&   \hdots&   \frac{1}{m_{\rm high}-nq(\psi)}
\end{pmatrix}.$$
The rational surface singularity becomes clear in that $\mathbf{Q^{-1}}$ will vanish when $q(\psi)=m/n$. Furthermore, near some singular surface $\psi_r$ where $m_r-n_rq(\psi_r)=0$, the resonant Fourier perturbation amplitude $\bar{\xi}_\psi|_{m_{r},n_{r}}$ takes on two asymptotic forms: a diverging energy solution and a finite energy solution. In the ideal stability problem, the diverging energy solution is set to zero \cite{glasser_direct_2016}.

\begin{figure}[b]
\includegraphics[width=1.0\linewidth]{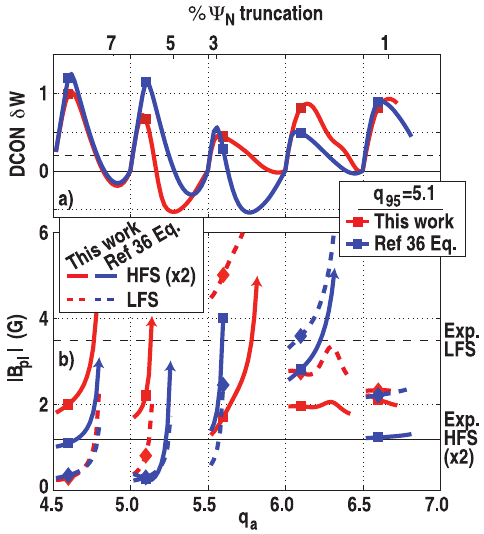}
\caption{\label{Paz} Figure reproduced with permission from \cite{paz-soldan_equilibrium_2016}, showing the $n=2$ edge plasma response of $\delta W$ and
$|B_{p1}|$ as equilibrium truncation q-value ($q_a$) is varied for two DIII-D equilibria. `Ref 36 Eq' refers to the second equilibrium, initially analysed in \cite{logan_dependence_2016}. $\psi_N$ refers to normalised poloidal flux. HFS and LFS refer to the high-field and low-field sides of the plasma respectively.}
\end{figure}

Despite the usefulness of magnetic coordinate formulations, they suffer a key setback in that the angle coordinates are undefined at the separatrix. This is because in a diverted tokamak plasma with a separatrix, the poloidal component of the magnetic field is zero at the x-point. The magnetic field lines do not close in the poloidal plane, and the periodic magnetic angle cannot be defined. Because of this, the computational domain must be truncated inside the separatrix in codes such as DCON \cite{glasser_direct_2016}, GPEC \cite{park_computation_2007,park_importance_2009,park_self-consistent_2017} and MARS \cite{liu_feedback_2000}. This would not be a problem\endnote{The computational domain must also be truncated approaching the magnetic axis. However as described in more detail in Appendix \ref{sec:Insensitive}, there is less sensitivity to this axis truncation point.} save for the sensitivity of stability codes to the edge truncation point. 

To illustrate this sensitivity of MHD stability calculations to edge truncation, consider the $\delta W$ value for a limited, low-pressure tokamak plasma. This will oscillate as $q_{eqdge}$ varies, becoming negative (and hence unstable) every time $n q_{eqdge}$ approaches a whole number from below \cite{wesson_hydromagnetic_1978,freidberg_ideal_2014}\endnote{This result was derived from analytic constructions of $\delta W$ in the large aspect ratio expansion, using $\xi_\psi$ test-functions increasingly localised at the plasma edge. For more information see Freidberg \cite{freidberg_ideal_2014} equation 11.234, figure 11.41 and surrounding text.}. This dependence of limited plasma stability on $q_{edge}$ is real, and has been experimentally observed as an external kink  \cite{mirnov_investigation_1971,makishima_simultaneous_1976,goodall_cine_1984,levesque_multimode_2013}.  
Oscillations with $q_{edge}$ are also present in the stability analysis of diverted configurations, despite not being `physical' as $q$ at the last closed flux surface is undefined. Examples of edge-oscillation in recent 3D field plasma response calculations can be found in \cite{paz-soldan_equilibrium_2016,li_modelling_2016,li_development_2021,xie_extension_2023} and \cite{bai_impact_2024}. The sharp dependence on $q_{edge}$ in these codes can at times obscure true physical mechanisms. 

The necessity of truncation in MHD stability calculations has resulted in diverted tokamaks with infinite rational surfaces being approximated by limited-like plasmas with a finite $q_{edge}$. If the truncation is chosen poorly, edge-localised instabilities can appear in the modelling where they do not in reality. To illustrate this, a calculation of the poloidal structure of the $q_{edge}$-dependent least-stable kink mode is shown in Fig. \ref{fig:Externalkink} for an ITER-like single-null DIII-D equilibrium (discharge 147131). If truncation is chosen inside of the $q=5$ surface, the model predicts that the plasma is unstable to an edge-localised kink mode that is not observed in experiments. Choosing the truncation to be just outside of the $q=5$ surface recovers the broader kink mode that is expected and consistent with plasma response validations of the GPEC/DCON model \cite{king_experimental_2015}.

We may implicitly assume that our stability calculations will approach a limit cycle as $q_{edge}$ goes to infinity in a diverted tokamak, and through this gain confidence about their predictions at high enough $q_{edge}$ (while insisting the truncation remain just outside of rather than just inside of a rational surface). However our calculations sometimes show oscillations that are not necessarily regular, such that they're not approaching a limit cycle on the domain where we can run them. This can be seen in Fig. \ref{Paz}, reproduced with permission from \cite{paz-soldan_equilibrium_2016}, which demonstrates the dependence of both the ideal perturbation energy $\delta W$ and the predicted $n=1$ edge poloidal error field amplitude $B_{p1}$ on the edge truncation point (both values computed in the GPEC/DCON code suite). Furthermore, some low $q95$ scenarios cannot even demonstrate a full oscillation in $n=1$ stability predictions before existing numerical methods of describing the equilibrium break down and the edge must be truncated, as shown in Fig. \ref{DCON_DW_calc}. Both of these scenarios demonstrate why edge truncation remains a problem in magnetic coordinate-dependent spectral stability codes.

There are currently two methods that decide where to truncate the equilibrium in GPEC/DCON. The first truncates at the largest value of $q=nk+\delta$ that is resolved by the equilibrium, where $k$ is an integer, $n$ is toroidal mode number, and $\delta$ is a user-input value usually between 0.1 and 0.3. Here $\delta$ is chosen to ensure $q_{edge}$ is always in the theoretically stable region of oscillations \cite{freidberg_ideal_2014}. This method implicitly assumes that evaluation further out is more `true,' however in practice predictions can become less reliable near the LCFS due to numerical issues. The second method is to scan $q_{edge}$ during the $\delta W$ calculation in DCON, and truncate the equilibrium at the point of maximum $\delta W$ in the resolved edge region. In practice this allows truncation at the largest $q_{edge}$ values while avoiding numerical instabilities, which anecdotally make $\delta W$ large and negative. However the drawback of method two is that the truncation point is being chosen based on what the user wants - stability. To illustrate a worst case scenario, this method could truncate an equilibrium at $q<5$ based on maximum $\delta W$, where it otherwise would be predicted unstable for $5<q<6$. Ultimately what is desired is one prediction for one diverted equilibrium. We will briefly outline the three barriers in place to achieving this.
\begin{table}[b]
\centering
\begin{tabular}{ | c | c | c | c |}
  \hline
  Name & $\alpha_p$ & $\alpha_B$ & $\alpha_R$ \\ \hline            
  Hamada & 0 & 0 & 0 \\ \hline      
  Boozer & 0 & 2 & 0 \\ \hline      
  Pest & 0 & 0 & 2 \\ \hline
  Equal-arc & 1 & 0 & 0 \\ \hline
\end{tabular}\caption{Parameters combinations that define poloidal magnetic angle via equations \ref{pol1} and \ref{pol2}. \label{tab:coordtypes}}
\end{table}

Resolving the edge truncation issue in diverted tokamak configurations requires overcoming three challenges:
\begin{enumerate}
    \item The computational domain becomes singular approaching the separatrix as one magnetic coordinate concentrates infinitely at the x-point. This effect is somewhat visible in figure 1 in \cite{park_spectral_2008}, where an x-point is present in the bottom left of the studied equilibrium. 
    To understand this singularity quantitatively, we must introduce the mathematical formulation of our magnetic angle coordinates. The generalised magnetic coordinate representation used in GPEC/DCON computes $\theta$ using a field-line tracing algorithm such that \begin{align}\label{pol1}
    \theta =\int \frac{\mathbf{B}\cdot \nabla\theta}{B_\eta} \mathrm{d}l_{\eta},
    \end{align}where $B_\eta=|\mathbf{B_p}\cdot \hat{\eta}|$ is the amplitude of the azimuthal component of the poloidal magnetic field $\mathbf{B_p}$, $\{r,\eta\}$ are polar coordinates in the poloidal plane originating at the magnetic axis, $\mathrm{d}l_{\eta}=r \hspace{0.5mm}\mathrm{d}\eta$ is the azimuthal component of the differential line element with units physical distance, and 
        \begin{align}\label{pol2}
    \mathbf{B} \cdot \nabla \theta = \frac{|\mathbf{B_p}|^{\alpha_p}|\mathbf{B}|^{\alpha_B}}{R^{\alpha_R}}
\end{align}

specifies the type of magnetic coordinate we are using. Common magnetic angle types are listed in table \ref{tab:coordtypes}.

Since there is no poloidal field at the x-point, the denominator term $B_\eta$ in the integral in Eq. \ref{pol1} approaches zero as the integrator gets close.   Hence for Boozer, Hamada, Pest and all coordinates where $\alpha_p < 1$, the $\theta$ integral diverges near x-points. An example calculation of this diverging integral is shown  in Fig. \ref{fig:eg_int}. The effect of this is that the magnetic angle $\theta$ increasingly concentrates in the vicinity of the x-point. Note since angle coordinates $\theta$ and $\zeta$ in GPEC/DCON are normalised to lie within ranges $[0,1)$, it is only the relative magnitude of the integral in Eq. \ref{pol1} that matters. In the equal-arc case, where $\alpha_p = 1$, poloidal angle $\theta$ does not diverge. However toroidal angle $\zeta$ diverges instead, since the two are related through the (diverging) $q$ profile \cite{park_spectral_2008}: $$\zeta=(\theta-\theta_0)q(\psi)$$
where $\theta_0$ is a constant on any given flux surface.

\item For a given $n$, the distance between rational surfaces goes to zero at the separatrix as $q$ goes approaches infinity. The current numerical methods used in DCON \cite{glasser_direct_2016} and resistive DCON \cite{glasser_computation_2016} to compute solutions either side of each rational surface singularity require a small distance between neighbouring rational surfaces to function. Ultimately some new treatment to address these infinitely bunched singularities is required if we are to overcome the edge truncation problem.

\begin{figure}[t]
\centering
\includegraphics[width=\linewidth]{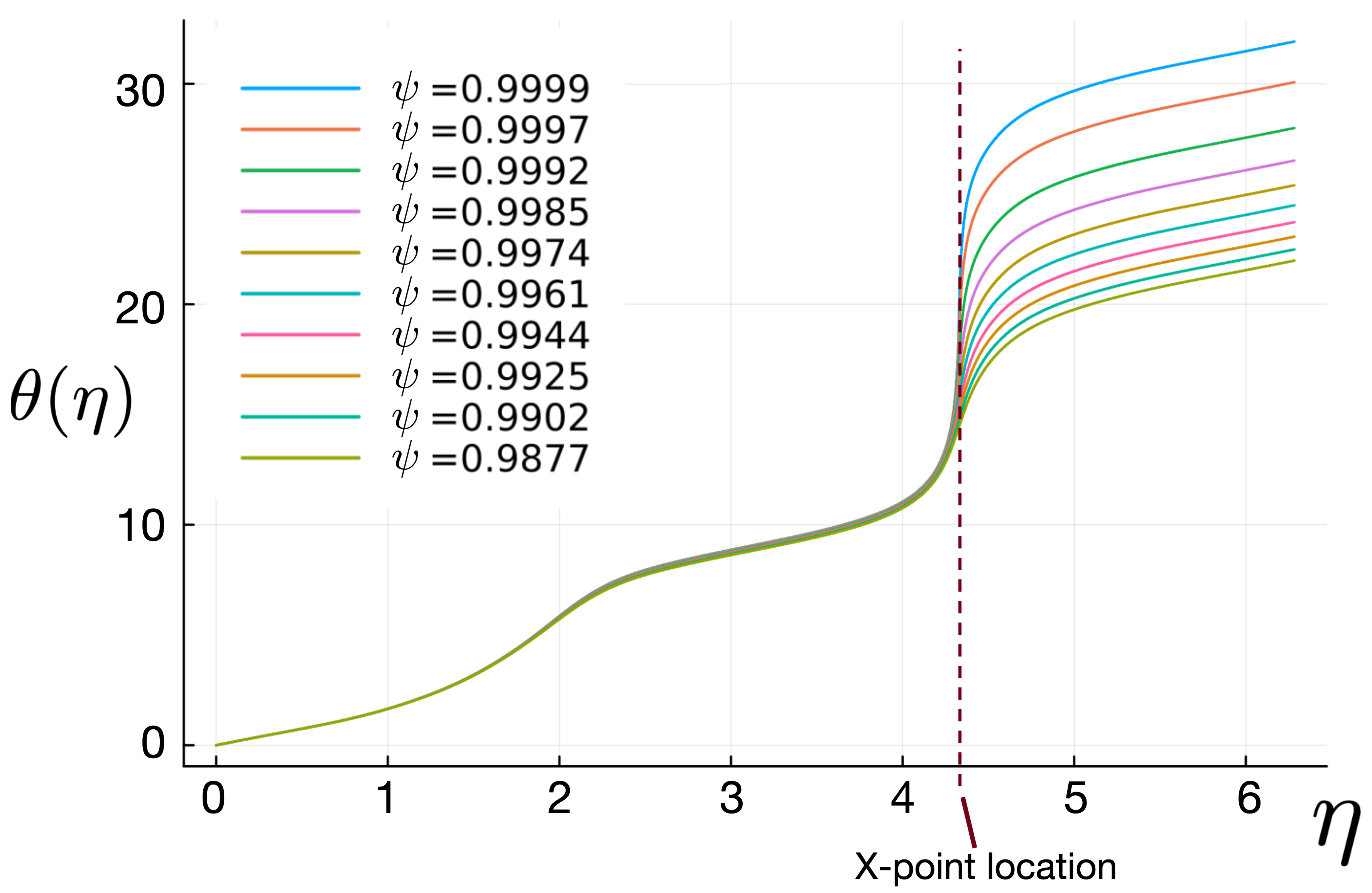}
\caption{\label{fig:eg_int} Example calculation of the field-line integral specified by equations \ref{pol1} and \ref{I4}, for Hamada $\theta$ coordinate, on flux surfaces approaching the separatrix of a single-null DIII-D equilibrium. $\psi$ here refers to normalised poloidal flux.}
\end{figure}

\item Aside from the rational surface singularities, there remains an underlying singularity in the $\bar{F}$ matrix (Eq. \ref{first_order_EL}) as $\psi$ approaches $\psi_{sep}$ \cite{glasser_direct_2016}. This second divergence may prevent the finite-energy component of the ideal perturbation solution going to zero even as the distance between neighbouring rational surfaces approaches 0 at the separatrix.

\end{enumerate}

The remains of this paper will address issue 1. For high resolution equilibria, issues 2 and 3 are currently the limiting factor during numerical calculation in GPEC/DCON. Note other attempts of dealing with edge truncation include artificially smoothing the x-point geometry to decouple $q_{edge}$ and $q_{95}$ effects \cite{yang_resistive_2019}, and in more recent work; imposing a cutoff value on the Jacobian to extract the divergent component of the PEST magnetic coordinate during stability calculations \cite{zheng_x-point_2025}. We note that while the latter method has greatly improved the smoothness of edge oscillations, a single-prediction has yet to be achieved for diverted stability calculations in spectral codes. In line with \cite{zheng_x-point_2025}, we believe physics beyond the ideal MHD model may inspire the correct solution to issues 2 and 3. For example, a truncation scheme based on resistive MHD physics has been proposed in \cite{turnbull_external_2016}.

\begin{figure*}
\centering
\includegraphics[width=0.8\textwidth]{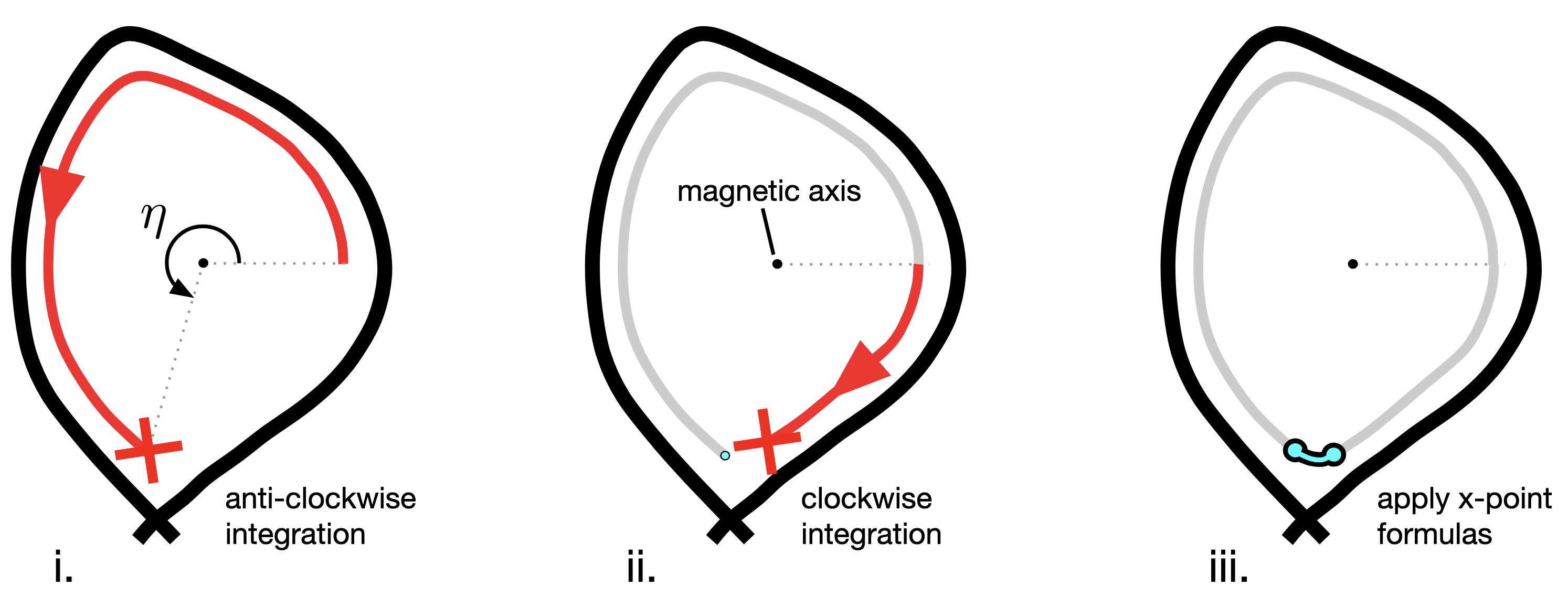}
\caption{\label{Intlogic} Illustration of numerical integrator logic in the vicinity of a single x-point, detailed in section \ref{sec:Interfacing}. The red crosses indicate the numerical integrator stopping due to $|\mathbf{B_p}|/|B_t|$ falling below some value $[B_p/B_t]_{tol}$ after attempting passes in the anticlockwise (i) and clockwise (ii) directions. The cyan line in panel iii represents the application of integral replacement formulas (\ref{Ia1}-\ref{Ia4}) for the missing arc. For double-null equilibria, the numerical integrator is re-initialised at $\eta=\pi$ for integration in both anti-clockwise and clockwise directions.}
\end{figure*}
\section{Calculation of magnetic coordinates asymptotically close to the separatrix}
In the following sections we describe our method for extending the computational domain in the GPEC/DCON code suite arbitrarily close to the separatrix in diverted equilibria. This involves the presentation of analytic formulas for the q-profile and magnetic poloidal coordinate $\theta$ in the divergent x-point region. Section \ref{sec:Interfacing} introduces the field line integrals that define these quantities, as well as the pre-existing numerical integration method used to compute them in the non-diverging region. Section \ref{sec:Derivation} describes the derivation of the analytic formulas, while their convergence properties are presented in section \ref{sec:results}. 

\subsection{Interfacing with field-line integrator}\label{sec:Interfacing}

In the GPEC/DCON code suite, an incoming equilibrium must be converted into magnetic coordinates specified by equations \ref{field_coords}, \ref{pol1} and \ref{pol2}. This process proceeds as follows for the common eqdsk file format \cite{lao_mhd_2005}: The poloidal flux is read in on an RZ-grid and converted into a 2D cubic spline, and then a field line integrator (LSODE \cite{hindmarsh_odepack_1983}) is initialised on flux surfaces of one's choosing. The integrator steps around in machine poloidal angle $\eta$, solving these integrals:
\begin{align}
    I_1(\eta) &= \int_0^\eta \frac{r }{B_\eta} \mathrm{d}\eta', \label{I1}
    \\
I_2(\eta) &= \int_0^\eta \frac{rB_r}{B_\eta} \mathrm{d}\eta', \label{I2}\\
I_3(\eta) &= \int_0^\eta \frac{r}{R^2B_\eta} \mathrm{d}\eta', \label{I3}\\
I_4(\eta) &= \int_0^\eta \frac{r }{B_\eta} \Bigl[\frac{|\mathbf{B_p}|^{\alpha_p}|\mathbf{B}|^{\alpha_B}}{R^{\alpha_R}}\Bigr] \mathrm{d}\eta', \label{I4}
\end{align}
which correspond to the following quantities:
\begin{align*}
    I_1&(2\pi) = \frac{1}{2\pi}\frac{dV}{d\psi},\\
I_2&(\eta) = r(\eta),\\
I_3&(2\pi) = 2\pi \frac{q(\psi)}{F(\psi)},\\
I_4&(\eta) = \theta(\eta).
\end{align*}
Note $V$ is plasma volume, $r$ is minor radius, $F$ is the flux function in the Grad-Shafranov equation \cite{freidberg_ideal_2014}, and flux coordinate $\psi$ is normalised poloidal flux. Algebraic combinations of $I_1,I_2,I_3$ and $I_4$ are written onto a grid of $\{\theta,\zeta,\psi\}$, and comprise all magnetic information needed for stability calculations. If the chosen flux surface is too close the the separatrix, then the integrator exceeds its maximum steps, or fails its error tolerance as it passes close to an x-point. To avoid this, new scripts have been developed that stop the field-line integrator when the ratio of the poloidal to toroidal field $|\mathbf{B_p}|/|B_t|$ drops below some input tolerance, and the missing arcs are replaced with analytic formulas. See Fig. \ref{Intlogic} for a visualisation of this process. When the $|\mathbf{B_p}|/|B_t|$ tolerance is exceeded, the ratio of the perpendicular distance from the separatrix divided by the distance from the x-point (`$d_{sep}/d_{X}$') is also checked. The replacement formulas are only applied if the `approach condition' \begin{equation}\label{dist_check}
    \frac{d_{sep}}{d_{X}}<\frac{1}{5},
    \end{equation}
 is met, to avoid cases where $|\mathbf{B_p}|/|B_t|$ is only exceeded at the point of closest approach. 

\subsection{Derivation of field-line integral replacement formulas}\label{sec:Derivation}

\begin{figure}[t]
\centering
\includegraphics[width=\linewidth]{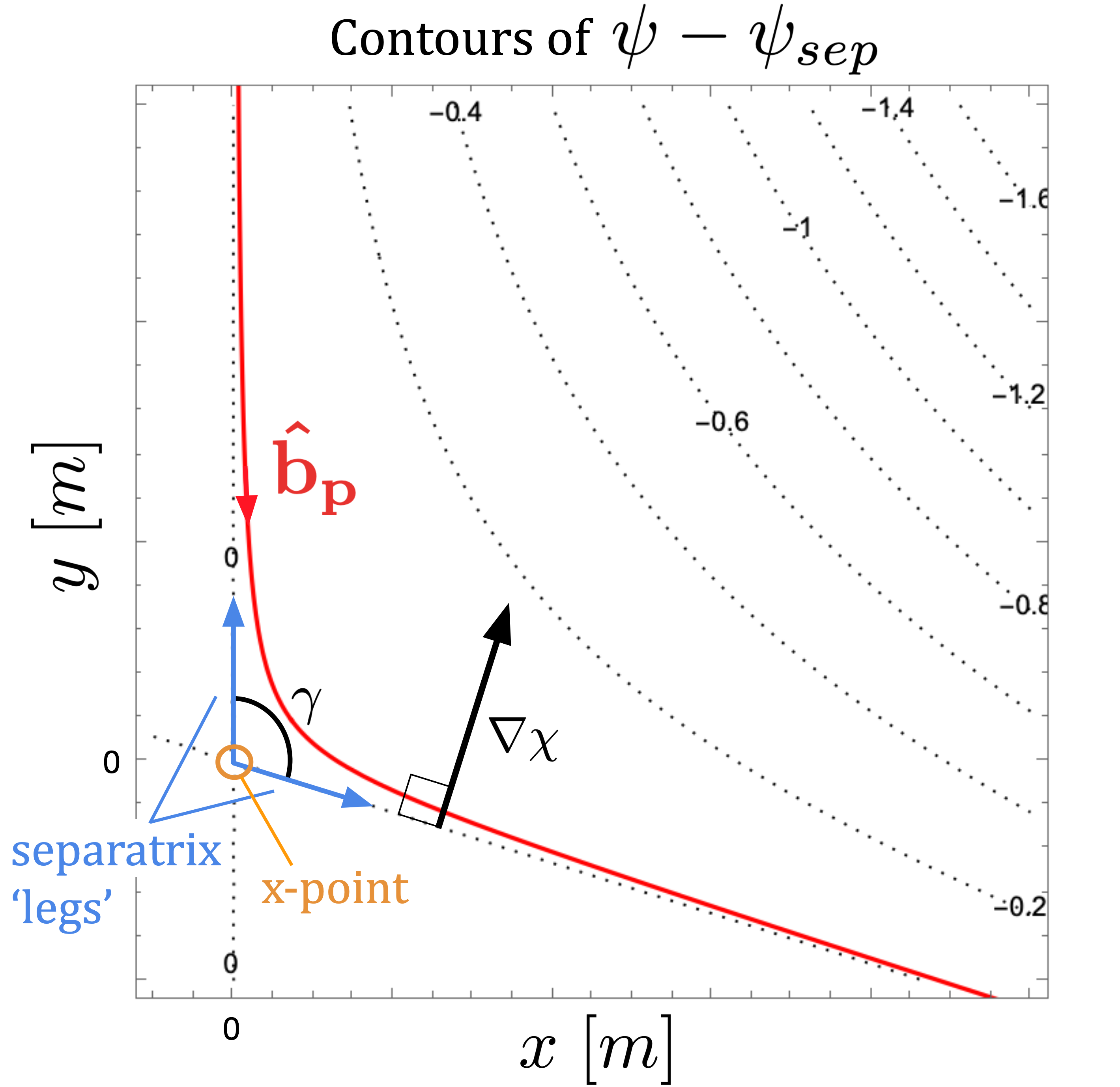}
\caption{\label{locaCoords} Illustrative plot of an x-point, in local $\{x,y\}$ coordinates defined in equations \ref{xcord} and \ref{ycord}. Dotted lines indicate contours of $\psi-\psi_{sep}$. The unit vector of the poloidal magnetic field $\color{red}\hat{b}_p$ aligns with these contours. Coordinate values $\gamma$ and $\nabla\chi$ have been included.}
\end{figure}

In this section we describe the formulation of analytic replacements for integrals $I_1,I_2,I_3$ and $I_4$, which are valid in the vicinity of the x-point. These formulas are based on three assumptions: that the two separatrix `legs' (labelled in Fig. \ref{locaCoords}) become straight infinitely close to the x-point, that $\psi$ near the x-point is smooth and can be described by a 2D Taylor expansion, and that $\psi$ has a non-zero lowest-order component at the separatrix. The first two assumptions are equivalent to requiring no infinitely sharp currents near the x-point, while the latter assumption should be true for all equilibria save special mathematically constructed cases. The ratios $|\mathbf{B_p}|/|B_t|$ and $B_\eta/|B_t|$ were also treated as small parameters. The procedure used to derive the field-line integral formulas is repeated here:

\begin{enumerate}
    \item A local x-y coordinate frame was constructed and aligned with one leg of the x-point:
    \begin{align}
    x &= R_{\rm local}\cos\vartheta  + Z_{\rm local}\sin \vartheta \label{xcord}\\
    y &= -R_{\rm local}\sin\vartheta + Z_{\rm local}\cos \vartheta. \label{ycord}
    \end{align}
    where $R_{\rm local} \equiv R-R_X$ and $Z_{\rm local} \equiv  Z-Z_X$ are local $R,Z$ coordinates centered on the x-point. $\vartheta$ is defined as the angle needed to rotate the $Z_{\rm local}$ axis in an anticlockwise direction until it aligns with the separatrix leg along which the integrator approaches the x-point during anti-clockwise integration (this is the $y$ axis in Fig. \ref{locaCoords}). 
    \item The angle between the separatrix legs $\gamma$ was used to define a spatial coordinate $\chi$:
        $$\chi=-\cos(\gamma)x+\sin(\gamma)y,$$
        such that $\nabla \chi$ is orthogonal to the other separatrix leg (see Fig. \ref{locaCoords}).
    Using $\chi$, the lowest order Taylor expansion of $\psi$ at the x-point simplifies to \begin{equation}\label{psi_tay}
    \psi=\psi_{sep} + c_{11}x\chi  
        \end{equation} where $c_{11}$ is the non-zero lowest order coefficient, and \begin{equation}\label{bp_local}
    \mathbf{B_p}=\frac{1}{R}\nabla \psi\times z.        \end{equation} Together equations \ref{psi_tay} and \ref{bp_local} give us the field-line trajectories $y(x)$: \begin{equation}
y(x) = \frac{C_0}{x} + x \cot(\gamma)  \label{ytraj}  
    \end{equation}

    where $C_0=x_1 y_1 - x_1^2 \cot(\gamma)$. This allows one to write the field-line integrals \ref{I1}-\ref{I4} in terms of x alone. The culmination of this process is the simplified\hspace{.3mm}x-point integral formula: \begin{align}\label{genform}&
        \int_{\text{sep}} g \frac{\mathrm{d}l_\eta}{B_\eta}\nonumber \\ =&\int_{\text{sep}} g \frac{\mathrm{d}l_{p}}{|\mathbf{B_p}|} = -\frac{\csc(\gamma)}{c_{11}} \int^{x_{2}}_{x_{1}}  \frac{R(x) g(x)}{x} \mathrm{d}x,\end{align}
        where $g$ is any equilibrium value being integrated. A guide for simple numerical methods to calculate $R_X,Z_X,\vartheta,\gamma$ and $c_{11}$ has been included in Appendix \ref{sec:method_of_calc}.
    \item Eq. \ref{genform} was applied to integrals \ref{I1}, \ref{I3} and \ref{I4} to give us the generalised x-point integral formulas in terms of $x$. To do this, the integrands $R$, $|\mathbf{B_p}|$ and $|\mathbf{B}|$ were expanded in terms of $x$ with ordering $\epsilon \sim x\sim y\sim  |\mathbf{B_p}|/|B_t|$, $R_X\sim |B_t|\sim O(1)$. Terms as small as $O(\epsilon)$ were retained. The completed formulas are:
    
\begin{widetext}
    \begin{equation}\label{Ia1}
           \int_{\text{sep}} \frac{\mathrm{d}l_\eta}{B_\eta} =-\frac{\csc(\gamma)}{c_{11}}  \Big[R_X \ln\frac{x_2}{x_1} + (x_2 - x_1)\Big(\cos\vartheta - \Big(1 - \frac{x_1}{x_2}\Big) \cot\gamma \hspace{0.5mm}\sin\vartheta\Big) - \Big(1 - \frac{x_1}{x_2}\Big) y_1 \sin\vartheta\Big],
    \end{equation}
       \begin{equation}
        \label{Ia3} 
            \int_{\text{sep}} \frac{1}{R^2}\frac{\mathrm{d}l_\eta}{B_\eta}=-\frac{\csc(\gamma)}{c_{11}}\Delta^{-\frac{1}{2}}\ln \frac{(x_2-\mathcal{R}_+)(x_1-\mathcal{R}_-)}{(x_1-\mathcal{R}_+)(x_2-\mathcal{R}_-)},
        \end{equation}
        where $\mathcal{R}_+$ and $\mathcal{R}_-$ are the positive and negative roots of the quadratic $xR(x)=Ax^2+Bx+C$, with $A=\cos (\vartheta )-\cot (\gamma ) \sin (\vartheta )$, $B=R_X$ and $C=x_1^2 \cot (\gamma ) \sin (\vartheta )-x_1 y_1 \sin (\vartheta )$. $\Delta=B^2-4AC$ is the usual quadratic determinant.
        \begin{align}
        \label{Ia4}
    \int_{\text{sep}} \frac{|\mathbf{B_p}|^{\alpha_p}|\mathbf{B}|^{\alpha_B}}{R^{\alpha_R}}\frac{\mathrm{d}l_\eta}{B_\eta}  &= -\csc(\gamma)|B_t|_X^{\alpha_B}\frac{|c_{11}^{\alpha_p}|}{c_{11}}R_X^{1-\alpha_R-\alpha_p} \times \nonumber\\ &\hspace{0.8cm} \Bigg[J_1 + \frac{(1-\alpha_R-\alpha_p)}{R_X}(\cos\vartheta  - \cot\gamma\sin \vartheta)J_2  - (1-\alpha_R-\alpha_p)\frac{\sin \vartheta}{R_X}(x_1 y_1 - x_1^2\cot\gamma)J_3\Bigg] ,
        \end{align}
    where
    \begin{align}
    J_1 &= \int^{x_{2}}_{x_{1}}  \frac{1}{x} \Bigg[x^2 + 2 \cos\gamma\hspace{1mm}  (x_1^2 \cos\gamma - x_1y_1 \sin\gamma) + 
  \frac{(x_1^2 \cos\gamma - x_1y_1 \sin\gamma)^2}   {x^2}\Bigg]^{\frac{\alpha_p}{2}}\mathrm{d}x,\\
    J_2 &= \int^{x_{2}}_{x_{1}} \Bigg[x^2 + 2 \cos\gamma\hspace{1mm}  (x_1^2 \cos\gamma - x_1y_1 \sin\gamma) + 
  \frac{(x_1^2 \cos\gamma - x_1y_1 \sin\gamma)^2}   {x^2}\Bigg]^{\frac{\alpha_p}{2}}\mathrm{d}x,\\
    J_3 &= \int^{x_{2}}_{x_{1}} \frac{1}{x^2}\Bigg[x^2 + 2 \cos\gamma\hspace{1mm}  (x_1^2 \cos\gamma - x_1y_1 \sin\gamma) + 
  \frac{(x_1^2 \cos\gamma - x_1y_1 \sin\gamma)^2}   {x^2}\Bigg]^{\frac{\alpha_p}{2}}\mathrm{d}x.
\end{align}
For the $\alpha_p=1$ case,
    \begin{align}
    J_1&= \text{Re}\Bigg[-\sqrt{\frac{d^2}{x^2}+2\hspace{.3mm}d \cos (\gamma )+x^2} \nonumber\\
    &\hspace{1.10cm}-2\hspace{.3mm}d\hspace{.3mm}x \sqrt{\frac{e^{i \gamma }}{d\hspace{.3mm}x^2}}\bigg((\sin (\gamma )+i \cos (\gamma )) E\bigg[i \sinh ^{-1}\bigg(\sqrt{\frac{e^{-i \gamma }}{d}} x\bigg)|e^{2 i \gamma }\bigg]-\sin (\gamma ) F\bigg[i \sinh ^{-1}\bigg(\sqrt{\frac{e^{-i \gamma }}{d}} x\bigg)|e^{2 i \gamma }\bigg]\bigg)\Bigg]^{x_2}_{x_1} \label{EE1}\\
    J_2&= \frac{1}{2}\Biggl[\sqrt{d^2+2\hspace{.3mm}d\hspace{.3mm}x^2 \cos \gamma+x^4}-d \tanh ^{-1}\bigg[\frac{d+x^2 \cos \gamma}{\sqrt{d^2+2\hspace{.3mm}d\hspace{.3mm}x^2 \cos \gamma+x^4}}\bigg]+d \cos \gamma \hspace{0.5mm}\tanh ^{-1}\bigg[\frac{d \cos \gamma+x^2}{\sqrt{d^2+2\hspace{.3mm}d\hspace{.3mm}x^2 \cos \gamma+x^4}}\bigg]\Biggr]^{x_2}_{x_1} \label{EE2}\\
    J_3&= \frac{1}{2}\Biggl[\tanh ^{-1}\bigg[\frac{d \cos \gamma+x^2}{\sqrt{d^2+2\hspace{.3mm}d\hspace{.3mm}x^2 \cos \gamma+x^4}}\bigg]-\cos \gamma \hspace{0.5mm}\tanh ^{-1}\bigg[\frac{d+x^2 \cos \gamma}{\sqrt{d^2+2\hspace{.3mm}d\hspace{.3mm}x^2 \cos \gamma+x^4}}\bigg]-x^{-2}\sqrt{d^2+2\hspace{.3mm}d\hspace{.3mm}x^2 \cos (\gamma )+x^4}\hspace{0.5mm}\Biggr]^{x_2}_{x_1} \label{EE3}
\end{align}

where $F$ and $E$ are incomplete elliptic integrals of the first and second kind respectively, and $d=(x_1^2 \cos\gamma - x_1y_1 \sin\gamma) < 0$. 
For Boozer, Hamada, Pest and all coordinates with $\alpha_p=0$, however, the $J_i$ integrals trivially simplify to 
$$J_1=\ln\Big(\frac{x_2}{x_1}\Big),\hspace{2mm}J_2=x_2-x_1,\hspace{2mm}\text{and }\hspace{2mm}J_3=\frac{1}{x_1}-\frac{1}{x_2}.$$
The integral replacement formulas for $\alpha_p=0$ are present in GPEC/DCON, while the $\alpha_p=1$ formulas were  computed in Mathematica for illustration here, and will be added to GPEC/DCON if demand is present. 
    
\item Equations \ref{Ia1}-\ref{Ia4} were converted back to $r,\eta$ to interface with the numerical field line integrator. This required solving for $x_2(\eta)$ by finding the intersection of the trajectory $y(x)$ (Eq. \ref{ytraj}) with a straight line from the magnetic axis ($x_{ax},y_{ax})$ specified by machine poloidal angle $\eta$:
    
    \begin{equation}
        x_2(\eta) = \frac{1}{2}\frac{y_{ax}-x_{ax}\tan(\eta-\vartheta)-\sqrt{[y_{ax}-x_{ax}\tan(\eta-\vartheta)]^2-4[\cot{\gamma}-\tan(\eta-\vartheta)][x_1 y_1 - x_1^2 \cot(\gamma)]}}{\cot{\gamma}-\tan(\eta-\vartheta)}\label{x2}
    \end{equation}

\begin{figure*}[t!]
\centering
\includegraphics[width=\textwidth]{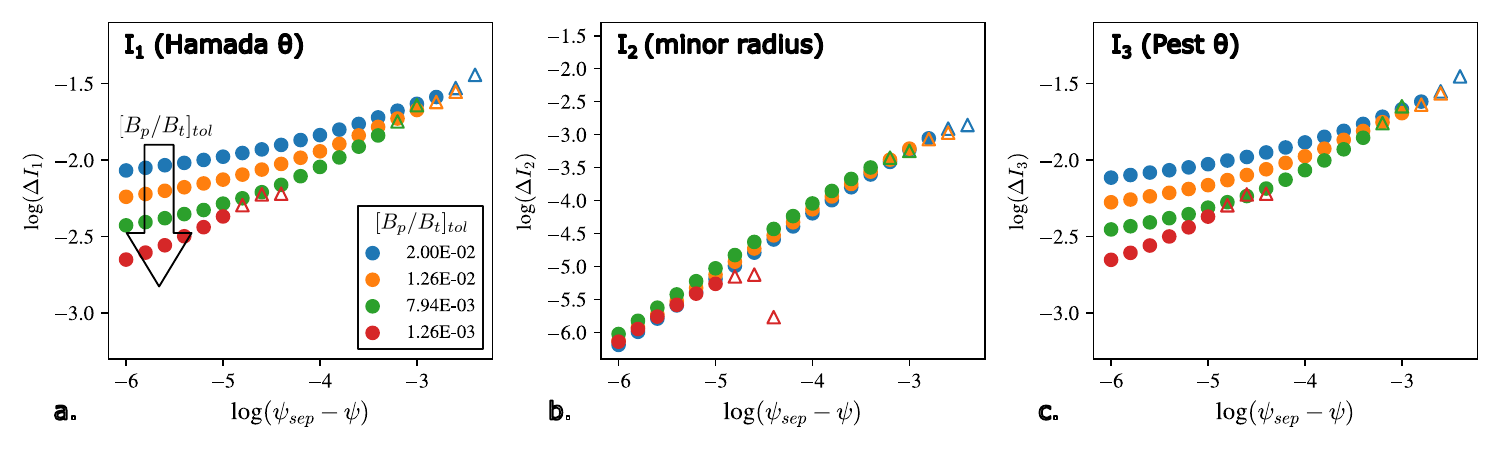}
\caption{\label{fig:Convg1} Relative error between the analytic expressions for $I_1$ (Eq. \ref{Ia1}), $I_2(x_2,y_2)$, and $I_3$ (Eq. \ref{Ia3}), and the numerical field-line integrator* across the x-point in the single-null DIII-D equilibrium shown in Fig. \ref{fig:Externalkink}. Distance between the flux-surface and the separatrix is varied along the x-axis. The colours represent different cases of input small parameter $[B_p/B_t]_{tol}$. The relative error formula used in the y-axis is given by Eq. \ref{Ierr1}. Triangles denote cases that failed to satisfy Eq. \ref{dist_check} and span extremely short integration arcs as a result.}
\end{figure*}
With $x_2(\eta)$ and $y_2=y(x_2)$ the final separatrix integral $I_2 = r$ (Eq. \ref{I2}) can be solved directly by converting $x_2,y_2$ back to $r,\eta$. 
\newpage
\end{widetext}
\end{enumerate}

\begin{figure}[b]
\centering
\includegraphics[width=0.75\linewidth]{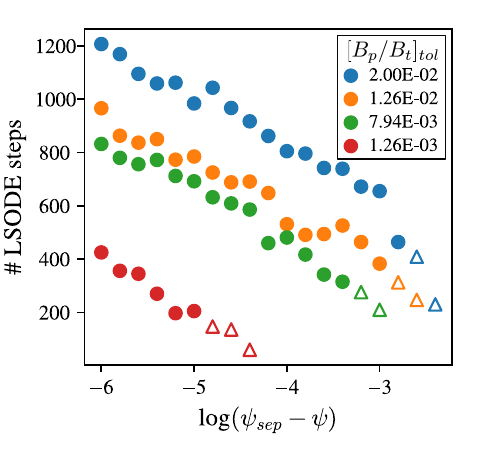}
\caption{\label{fig:NumLSODE_STEPS} Number of steps taken by the LSODE numerical field-line integrator when calculating integrals $I_1$-$I_3$, as used in the comparison cases in Fig. \ref{fig:Convg1}.}
\end{figure}

\section{Results}\label{sec:results}

\begin{figure*}[t]
\centering
\includegraphics[width=0.6666\textwidth]{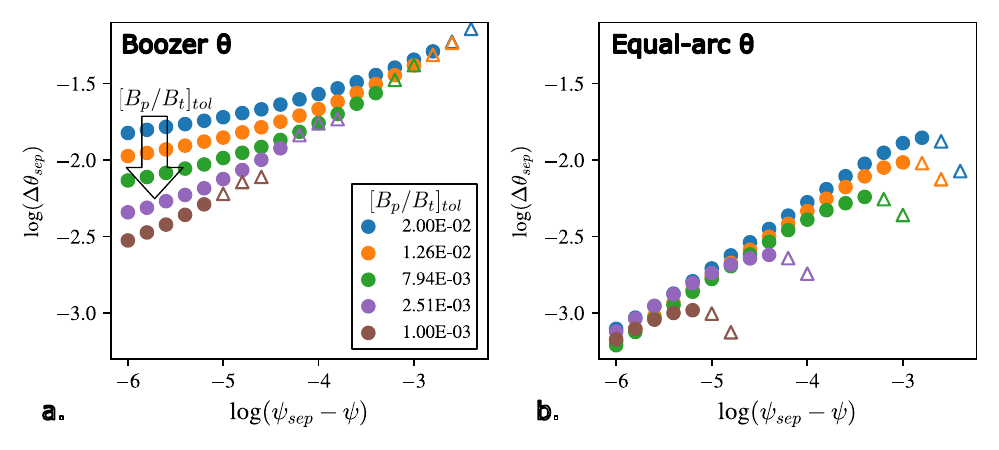}
\caption{\label{fig:Convg2} Relative error between the analytic $\theta$ formula given by Eq. \ref{Ia4} and numerical field-line integrator* across the x-point in the single-null DIII-D equilibrium shown in Fig. \ref{fig:Externalkink}. Panels a. and b. show Boozer and equal-arc poloidal angles respectively, with the equal-arc case calculated using formulas \ref{EE1}-\ref{EE3}. Distance between the flux-surface and the separatrix is varied along the x-axis. The colours represent different cases of input small parameter $[B_p/B_t]_{tol}$, and the relative error formula $\Delta \theta_{sep}$ is specified by Eq. \ref{Ierr1}. Triangles denote cases that failed to satisfy Eq. \ref{dist_check} and span extremely short integration arcs as a result.}
\end{figure*}

Convergence tests of the new integral replacement formulas are presented in Figs. \ref{fig:Convg1} and \ref{fig:Convg2}. These results were computed using the same $128\times128$ eqdsk equilibrium file from DIII-D discharge 147131 that is shown in Fig. \ref{fig:Externalkink}. In all cases the analytic formulas were bench-marked against the original numerical integrator running (slowly) at increased maximum total steps and maximum resolution, in a region* where $\psi$ was close enough to $\psi_{edge}$ that the integrals were beginning to diverge, but not so close that the numerical integrator's step-wise relative error tolerance of $10^{-14}$ was exceeded (a demonstration the numerical integrator breaking down is included in Appendix \ref{sec:NumIntFail}). The y-axis of all cases represents the relative difference between the analytic formula and numerical integrator over the $\eta$-arc specified by $[B_p/B_t]_{tol}$ (visualised in Fig. \ref{Intlogic}iii). The relative error formula used is
\begin{equation}
    \Delta I_i = \frac{|I_{i,n}-I_{i,a}|}{I_{i,n}}  \text{ for i}=\{1,4\}\label{Ierr1}\\
\end{equation}
where $\Delta \theta_{sep}\equiv \Delta I_4$, and subscripts $_a$ and $_n$ denote the analytic and numerical calculations respectively.\\

In general the relative accuracy of the analytic formulas increases for flux surfaces closer to the separatrix. For the non-diverging integrals of minor radius $(I_2)$ and equal-arc $\theta$, shown in Figs. \ref{fig:Convg1}b and \ref{fig:Convg2}b  respectively, this relationship is a straight line in the log-log plot, indicating power-like convergence, with little dependence on $[B_p/B_T]_{tol}$. For $\theta$ angles with $\alpha_p=0$, including Hamada (\ref{fig:Convg1}a), Pest (\ref{fig:Convg1}b), and Boozer coordinates (\ref{fig:Convg2}a), convergence approaches a straight line only for increasingly small values of $[B_p/B_T]_{tol}$. The magnitude of the relative error also gets smaller with smaller $[B_p/B_T]_{tol}$. This is to be expected: the accuracy of the integral formulas is predicated on the assumptions laid out in section \ref{sec:Derivation}, including using a lowest-order Taylor expansion in $\psi$, and treating $x$, $y$ and $|\mathbf{B_p}|/|\mathbf{B}|$ as small parameters. All these assumptions and approximations become increasingly true closer to the x-point, and the proximity of the path of integration to the x-point is controlled directly by $\psi_{sep}-\psi$, and indirectly by $[B_p/B_T]_{tol}$. The smallness of $|\mathbf{B_p}|/|\mathbf{B}|$ is also directly limited by $[B_p/B_T]_{tol}$.

The approach condition (Eq. \ref{dist_check}) when failed appears as a triangle in Figs. \ref{fig:Convg1} and \ref{fig:Convg2}. These cases manifest as extremely short integration arcs, which cause a large reduction in the relative error of the non-divergent integrals (\ref{fig:Convg1}b and \ref{fig:Convg2}b) because the denominator in Eq. \ref{Ierr1} shrinks faster than the numerator. These short-arc cases have little affect on the convergence property of the diverging integrals, however, since most of the integration value occurs at the point of closest approach to the x-point.

Finally, the number of steps taken by the LSODE integrator for each diverging integral is shown in \ref{fig:NumLSODE_STEPS}, where each point in that graph is equivalent to a single line calculated using the analytic formulas.

\section{Conclusion}

Spectral stability codes continue to be hindered by an inherent singularity in their magnetic coordinate formulation, that appears at the separatrix in diverted plasma configurations. In this paper we present an analytical description of this divergence for generalised magnetic coordinates, that can be used to extend the computational domain asymptotically close to the x-point. These integral formulas demonstrate power-like convergence with respect to $\psi_{sep}-\psi$, and are controlled by a single small-parameter input $|\mathbf{B_p}|/|B_t|$ that determines their accuracy. Unlike the numerical integrator they replace, these formulas retain the correct log-linear trend when approaching arbitrarily close to the separatrix, without accruing increased computational cost. 
The description of magnetic coordinates afforded by these integrals addresses the first of three challenges that currently comprise the edge-truncation problem in spectral MHD stability codes. We leave challenges two and three, which require new asymptotic procedures to deal with the infinite build-up of rational surface singularities at the edge, not to mention an additional underlying separatrix singularity, to future work.

\begin{figure*}[t]
\includegraphics[width=0.9\linewidth]{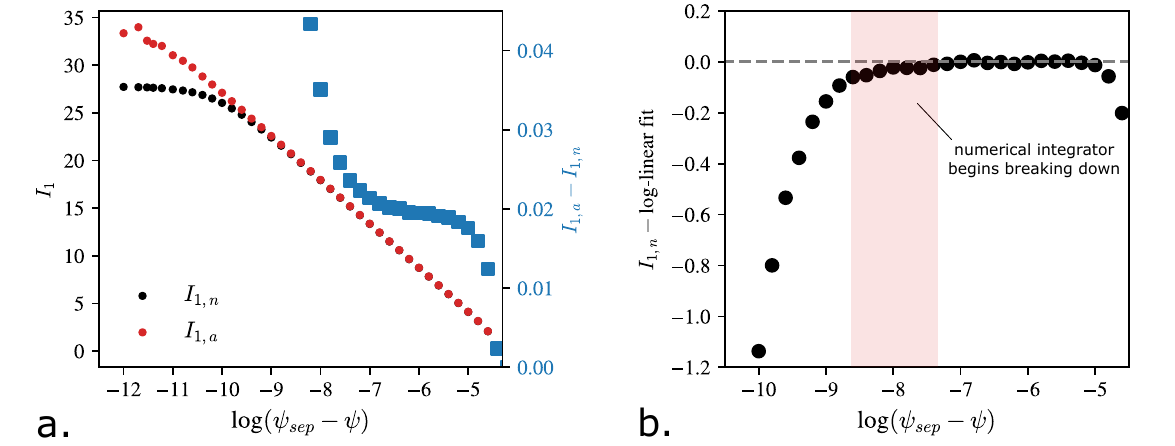}
\caption{\label{fig:I1_trends} Demonstration of numerical integrator breaking down at the far edge during calculation of $I_1$ as the distance from the separatrix is varied. Panel a plots $I_1$ for both numerical $(I_{1,n})$ and analytic $(I_{1,a})$ cases, as well as the difference between them. The divergence of $I_{1,a}$ from the log-linear trend in the top left corner is due to finite-precision Newton method errors when calculating $x_1$ for values less than $10^{-10}$, explained in Appendix \ref{sec:method_of_calc}. Panel b shows the deviation of $I_{1,n}$ from the log-linear fit$^*$ specified by Eq. \ref{equ test}. $[B_p/B_t]_{tol}$ was set to $1.26$E-03 for both these calculations.}
\end{figure*}

\theendnotes


\section*{Appendix}
\appendix

\subsection{Quantifying the failure point of the numerical integrator}\label{sec:NumIntFail}

We include Fig. \ref{fig:I1_trends} to illustrate the break-down of the numerical field line integrator as the flux surface of interest approaches the separatrix.  $I_1$ is calculated in Fig. \ref{fig:I1_trends} using both the numerical integrator and analytic formula over a large range in $\psi$. In panel a, both cases appear to follow a log-linear trend until $\psi-\psi_{sep}$ approaches $10^{-9}$, where the numerical integrator starts to (incorrectly) plateau. However the numerical integrator begins to introduce error earlier, around $\psi-\psi_{sep}\sim 10^{-7.5}$. Indirect evidence of this can be found in the difference between the numerical and analytic calculations in Fig. \ref{fig:I1_trends}a, which stays nearly constant for $\psi-\psi_{sep}\in (10^{-7},10^{-5})$ before rapidly increasing  for $\psi-\psi_{sep}<10^{-7}$. In Fig. \ref{fig:I1_trends}b a straight-line fit was applied to the numerical calculation,
\begin{equation}\label{equ test}
    \text{log-linear fit}= -4.61\times\log(\psi_{sep}-\psi)-18.9,
\end{equation}
and subtracted from $I_{1,n}$. Through this we can directly observe the numerical integrator diverging from the (correct) log-linear trend around $\psi-\psi_{sep}\sim 10^{-7.5}$. In accordance with these results, the domain $\psi-\psi_{sep} \geq 10^{-6}$ was used for quantitative convergence studies of the analytic formulas in section \ref{sec:results}.

\subsection{Insensitivity of GPEC/DCON to truncation near the axis}\label{sec:Insensitive}

The computational domain in magnetic coordinate-dependent spectral stability codes must also be truncated approaching the magnetic axis, to avoid the natural singularity in poloidal angle $\theta$ that occurs at $\psi=0$. However unlike at the plasma edge, codes such as DCON \cite{glasser_direct_2016} are insensitive to the point of truncation in this case. Fig. \ref{fig:InnerTrunc} illustrates this by showing how the minimum-$\delta W$ plasma displacements only experience minor change for variations in lower-bound truncation point $(\psi_{low})$ up to 10\% of the total plasma $\psi$. In comparison, changing the upper-bound truncation point $\psi_{edge}$ by only 0.5\% can greatly change the structure of the minimum energy perturbations, as shown in Fig. \ref{fig:Externalkink}.

\subsection{Methods for computing $\boldsymbol{\{R_{local},Z_{local},\vartheta,\gamma,c_{11}\}}$}\label{sec:method_of_calc}

To aid in the application of the integrals presented in this paper, we outline below some useful methods for computing the terms that they depend upon. The precise location of the x-point  $R_X,Z_X$ can be found with a 2D Newton method iterating to within some tolerance in $dR$, $dZ$:
    \begin{align*}
        \mathrm{det}&=\begin{vmatrix}
   \partial_R B_R &   \partial_Z B_R \\
   \partial_R B_Z & \partial_Z B_Z    \\
\end{vmatrix}_{R_i,Z_i}\\
        dR&=\frac{B_Z\partial_Z B_R -B_R\partial_Z B_Z}{\mathrm{det}}\Bigg|_{R_i,Z_i}\\
        dZ&=\frac{B_R\partial_R B_Z-B_Z\partial_R B_R}{\mathrm{det}}\Bigg|_{R_i,Z_i}\\
        R_{i+1}&=R_{i}+dR\\
        Z_{i+1}&=Z_{i}+dZ
    \end{align*}
%
With $R_X,Z_X$ found, $R_{\rm local} \equiv R-R_X$ and $Z_{\rm local} \equiv  Z-Z_X$ can be defined, as well as a local polar coordinate system $\nu,\rho$:
\begin{figure}[b]
\includegraphics[width=\linewidth]{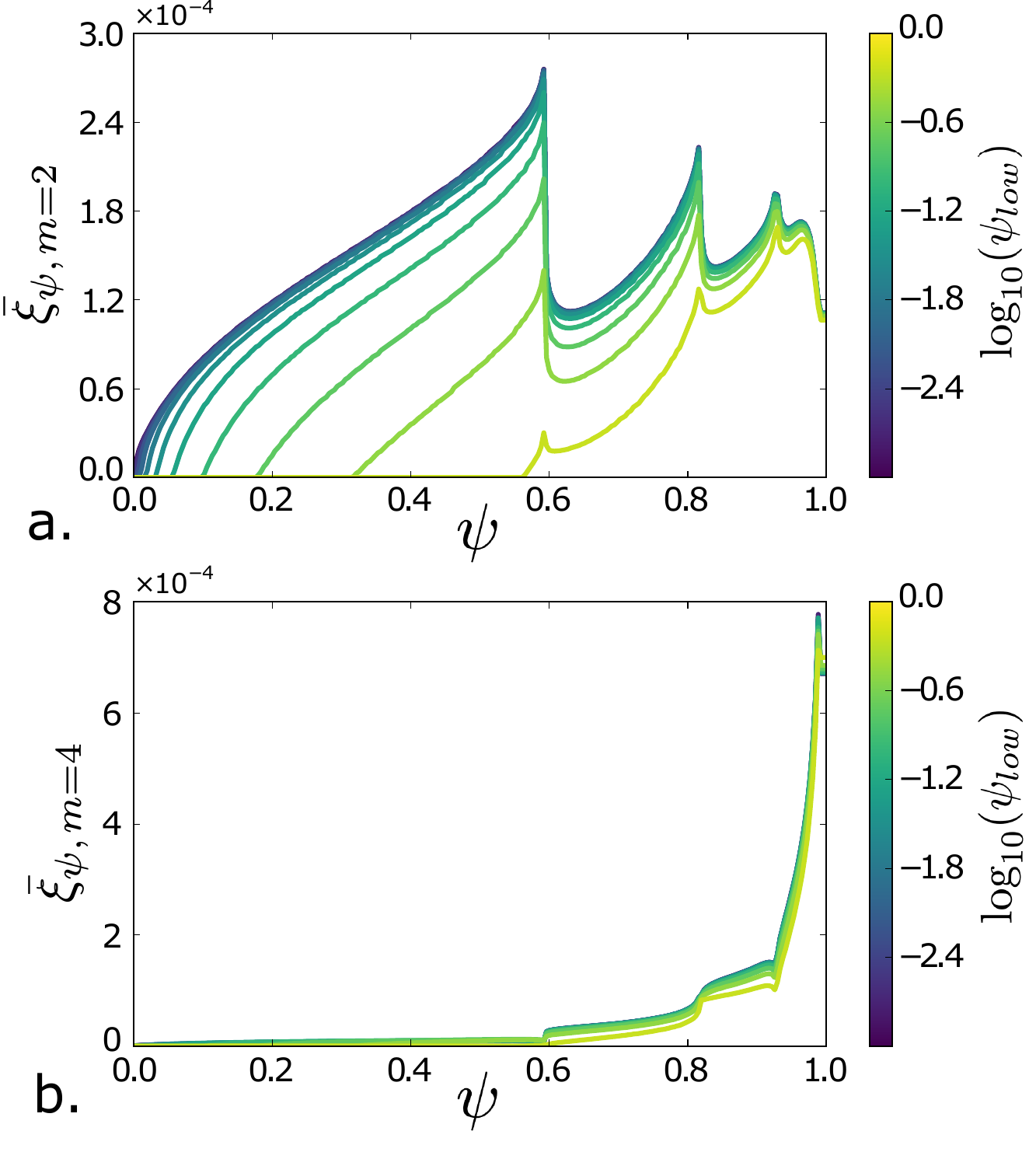}
\caption{\label{fig:InnerTrunc} Plot of the $n=1$ plasma displacement $\bar{\xi}_\psi$ as a function of normalised poloidal flux $\psi$, as inner truncation point $\psi_{low}$ is varied. Panels a  and b plot  the $m=2$ and $m=4$ poloidal components respectively.}
\end{figure}

\begin{align}
\nu&=\arctan \Big(\frac{Z_{\rm local}}{R_{\rm local}}\Big)\nonumber \\
\rho&=\sqrt{R_{\rm local}^2+Z_{\rm local}^2} \label{simpAng}
\end{align}

The $\nu$-angles of the x-point legs, which can be used to calculate $\vartheta$ and $\gamma$, can be found as follows; the numerical integrator, when it exceeds $[B_p/B_t]_{tol}$, will provide points $(r_1,\eta_1)$ and $(r_2,\eta_2)$ which serve as integration bounds for the analytic formulas. At each integration bound $\eta_i$, a Newton-method in radius can be applied to find the separatrix radius $r_{sep}(\eta_i)$, using formulas
\begin{align*}
    dr &= \frac{\psi_{sep}-\psi}{\cos(\eta_i)\partial_R \psi +\sin(\eta_i)\partial_Z \psi }\Bigg|_{r_i,\eta_i},\\
    r_{i+1}&=r_i+dr.
\end{align*}
Once the points $r_{sep}(\eta_1),\eta_1$ and $r_{sep}(\eta_2),\eta_2$ have been found, they can be easily converted to $\nu_1,\rho_1$, and $\nu_2,\rho_2$ via $R_{\rm local},Z_{\rm local}$ coordinates. Angles $\vartheta$ and $\gamma$ are then given by: \begin{align}
    \vartheta&=\nu_1-\pi/2,\\
    \gamma &=\nu_1-\nu_2.
\end{align}
Note $\nu_1$ and $\nu_2$ must be computed as above, through the straight lines defined by $\eta_1$ and $\eta_2$. This is to ensure $x_1$, which precisely controls the divergent behaviour of the x-point formulas, is correctly defined. Furthermore the error tolerances of the various Newton methods outlined above should be adjusted accordingly to calculating $x_1$ values to machine precision \cite{hansen_tokamaker_2024}.

Finally the lowest-order non-zero Taylor coefficient component of $\psi$, $c_{11}$ can found by the following method:
\begin{enumerate}
    \item As some small local radius $\rho_c$, use a bisection algorithm (or Newton method) to find the angle $\nu_{c}$ between the x-point legs at which $\mathbf{B_p}\cdot \hat{\rho} = \cos(\nu) B_{R}+\sin(\nu) B_Z$ goes to zero.
    \item Should $\psi$ be exactly described by its lowest-order Taylor expansion, as in Eq. \ref{psi_tay}, $\nu_{c}$ will lie precisely halfway between the x-point legs. The difference between $\nu_{c}$ and the half-way point is measured in GPEC/DCON to ensure the lowest-order Taylor expansion approximation is accurate. 
    \item After $\nu_{c}$ and $\rho_c$ are converted to $x,\chi$ coordinates via their definitions in section \ref{sec:Derivation}, and the point $(x_c,\chi_c)$ is found, $c_{11}$ can be computed with the formula \begin{equation}
        c_{11}=\frac{\psi_X-\psi_c}{x_c \chi_c}
    \end{equation}
    where $\psi_X$ and $\psi_c$ are the values of $\psi$ at the x-point and $(\nu_{c}, \rho_c)$, respectively.
\end{enumerate}

\section*{Acknowledgements}

Work supported by Commonwealth Fusion Systems. This material is based upon work supported by the U.S. Department of Energy, Office of Science, Office of Fusion Energy Sciences, using the DIII-D National Fusion Facility, a DOE Office of Science user facility, under Awards DE-FC02-04ER54698, DE-SC0014264, DE-SC0022270, and DE-SC0024898. 

\textbf{Disclaimer:} This report was prepared as an account of work sponsored by an agency of the United States Government. Neither the United States Government nor any agency thereof, nor any of their employees, makes any warranty, express or implied, or assumes any legal liability or responsibility for the accuracy, completeness, or usefulness of any information, apparatus, product, or process disclosed, or represents that its use would not infringe privately owned rights. Reference herein to any specific commercial product, process, or service by trade name, trademark, manufacturer, or otherwise does not necessarily constitute or imply its endorsement, recommendation, or favoring by the United States Government or any agency thereof. The views and opinions of authors expressed herein do not necessarily state or reflect those of the United States Government or any agency thereof.

\section*{Bibliography}

\bibliography{MyLibrary4}

\begin{thebibliography}{31}%
\makeatletter
\providecommand \@ifxundefined [1]{%
 \@ifx{#1\undefined}
}%
\providecommand \@ifnum [1]{%
 \ifnum #1\expandafter \@firstoftwo
 \else \expandafter \@secondoftwo
 \fi
}%
\providecommand \@ifx [1]{%
 \ifx #1\expandafter \@firstoftwo
 \else \expandafter \@secondoftwo
 \fi
}%
\providecommand \natexlab [1]{#1}%
\providecommand \enquote  [1]{``#1''}%
\providecommand \bibnamefont  [1]{#1}%
\providecommand \bibfnamefont [1]{#1}%
\providecommand \citenamefont [1]{#1}%
\providecommand \href@noop [0]{\@secondoftwo}%
\providecommand \href [0]{\begingroup \@sanitize@url \@href}%
\providecommand \@href[1]{\@@startlink{#1}\@@href}%
\providecommand \@@href[1]{\endgroup#1\@@endlink}%
\providecommand \@sanitize@url [0]{\catcode `\\12\catcode `\$12\catcode `\&12\catcode `\#12\catcode `\^12\catcode `\_12\catcode `\%12\relax}%
\providecommand \@@startlink[1]{}%
\providecommand \@@endlink[0]{}%
\providecommand \url  [0]{\begingroup\@sanitize@url \@url }%
\providecommand \@url [1]{\endgroup\@href {#1}{\urlprefix }}%
\providecommand \urlprefix  [0]{URL }%
\providecommand \Eprint [0]{\href }%
\providecommand \doibase [0]{http://dx.doi.org/}%
\providecommand \selectlanguage [0]{\@gobble}%
\providecommand \bibinfo  [0]{\@secondoftwo}%
\providecommand \bibfield  [0]{\@secondoftwo}%
\providecommand \translation [1]{[#1]}%
\providecommand \BibitemOpen [0]{}%
\providecommand \bibitemStop [0]{}%
\providecommand \bibitemNoStop [0]{.\EOS\space}%
\providecommand \EOS [0]{\spacefactor3000\relax}%
\providecommand \BibitemShut  [1]{\csname bibitem#1\endcsname}%
\let\auto@bib@innerbib\@empty
\bibitem [{\citenamefont {Bai}\ \emph {et~al.}(2024)\citenamefont {Bai}, \citenamefont {Loarte}, \citenamefont {Liu}, \citenamefont {Pinches}, \citenamefont {Koechl}, \citenamefont {Li}, \citenamefont {Dubrov},\ and\ \citenamefont {Gribov}}]{bai_impact_2024}%
  \BibitemOpen
  \bibfield  {author} {\bibinfo {author} {\bibnamefont {Bai}, \bibfnamefont {X.}}, \bibinfo {author} {\bibnamefont {Loarte}, \bibfnamefont {A.}}, \bibinfo {author} {\bibnamefont {Liu}, \bibfnamefont {Y.~Q.}}, \bibinfo {author} {\bibnamefont {Pinches}, \bibfnamefont {S.~D.}}, \bibinfo {author} {\bibnamefont {Koechl}, \bibfnamefont {F.}}, \bibinfo {author} {\bibnamefont {Li}, \bibfnamefont {L.}}, \bibinfo {author} {\bibnamefont {Dubrov}, \bibfnamefont {M.}}, \ and\ \bibinfo {author} {\bibnamefont {Gribov}, \bibfnamefont {Y.}},\ }\bibfield  {title} {\enquote {\bibinfo {title} {Impact of increasing plasma-wall gap on plasma response to {RMP} fields in {ITER}},}\ }\href {\doibase 10.1088/1361-6587/ad3aa0} {\bibfield  {journal} {\bibinfo  {journal} {Plasma Physics and Controlled Fusion}\ }\textbf {\bibinfo {volume} {66}},\ \bibinfo {pages} {055017} (\bibinfo {year} {2024})}\BibitemShut {NoStop}%
\bibitem [{\citenamefont {Bialek}\ \emph {et~al.}(2001)\citenamefont {Bialek}, \citenamefont {Boozer}, \citenamefont {Mauel},\ and\ \citenamefont {Navratil}}]{bialek_modeling_2001}%
  \BibitemOpen
  \bibfield  {author} {\bibinfo {author} {\bibnamefont {Bialek}, \bibfnamefont {J.}}, \bibinfo {author} {\bibnamefont {Boozer}, \bibfnamefont {A.~H.}}, \bibinfo {author} {\bibnamefont {Mauel}, \bibfnamefont {M.~E.}}, \ and\ \bibinfo {author} {\bibnamefont {Navratil}, \bibfnamefont {G.~A.}},\ }\bibfield  {title} {\enquote {\bibinfo {title} {Modeling of active control of external magnetohydrodynamic instabilities},}\ }\href {\doibase 10.1063/1.1362532} {\bibfield  {journal} {\bibinfo  {journal} {Physics of Plasmas}\ }\textbf {\bibinfo {volume} {8}},\ \bibinfo {pages} {2170--2180} (\bibinfo {year} {2001})}\BibitemShut {NoStop}%
\bibitem [{\citenamefont {Freidberg}(2014)}]{freidberg_ideal_2014}%
  \BibitemOpen
  \bibfield  {author} {\bibinfo {author} {\bibnamefont {Freidberg}, \bibfnamefont {J.~P.}},\ }\href@noop {} {\emph {\bibinfo {title} {Ideal {MHD}}}}\ (\bibinfo  {publisher} {Cambridge University Press},\ \bibinfo {year} {2014})\BibitemShut {NoStop}%
\bibitem [{\citenamefont {Glasser}(2016)}]{glasser_direct_2016}%
  \BibitemOpen
  \bibfield  {author} {\bibinfo {author} {\bibnamefont {Glasser}, \bibfnamefont {A.~H.}},\ }\bibfield  {title} {\enquote {\bibinfo {title} {The direct criterion of {Newcomb} for the ideal {MHD} stability of an axisymmetric toroidal plasma},}\ }\href {\doibase 10.1063/1.4958328} {\bibfield  {journal} {\bibinfo  {journal} {Physics of Plasmas}\ }\textbf {\bibinfo {volume} {23}},\ \bibinfo {pages} {072505} (\bibinfo {year} {2016})}\BibitemShut {NoStop}%
\bibitem [{\citenamefont {Glasser}, \citenamefont {Wang},\ and\ \citenamefont {Park}(2016)}]{glasser_computation_2016}%
  \BibitemOpen
  \bibfield  {author} {\bibinfo {author} {\bibnamefont {Glasser}, \bibfnamefont {A.~H.}}, \bibinfo {author} {\bibnamefont {Wang}, \bibfnamefont {Z.~R.}}, \ and\ \bibinfo {author} {\bibnamefont {Park}, \bibfnamefont {J.-K.}},\ }\bibfield  {title} {\enquote {\bibinfo {title} {Computation of resistive instabilities by matched asymptotic expansions},}\ }\href {\doibase 10.1063/1.4967862} {\bibfield  {journal} {\bibinfo  {journal} {Physics of Plasmas}\ }\textbf {\bibinfo {volume} {23}},\ \bibinfo {pages} {112506} (\bibinfo {year} {2016})}\BibitemShut {NoStop}%
\bibitem [{\citenamefont {Glasser}\ and\ \citenamefont {Kolemen}(2018)}]{glasser_robust_2018}%
  \BibitemOpen
  \bibfield  {author} {\bibinfo {author} {\bibnamefont {Glasser}, \bibfnamefont {A.~S.}}\ and\ \bibinfo {author} {\bibnamefont {Kolemen}, \bibfnamefont {E.}},\ }\bibfield  {title} {\enquote {\bibinfo {title} {A robust solution for the resistive {MHD} toroidal $\delta'$ matrix in near real-time},}\ }\href {\doibase 10.1063/1.5029477} {\bibfield  {journal} {\bibinfo  {journal} {Physics of Plasmas}\ }\textbf {\bibinfo {volume} {25}},\ \bibinfo {pages} {082502} (\bibinfo {year} {2018})}\BibitemShut {NoStop}%
\bibitem [{\citenamefont {Glasser}, \citenamefont {Kolemen},\ and\ \citenamefont {Glasser}(2018)}]{glasser_riccati_2018}%
  \BibitemOpen
  \bibfield  {author} {\bibinfo {author} {\bibnamefont {Glasser}, \bibfnamefont {A.~S.}}, \bibinfo {author} {\bibnamefont {Kolemen}, \bibfnamefont {E.}}, \ and\ \bibinfo {author} {\bibnamefont {Glasser}, \bibfnamefont {A.~H.}},\ }\bibfield  {title} {\enquote {\bibinfo {title} {A {Riccati} solution for the ideal {MHD} plasma response with applications to real-time stability control},}\ }\href {\doibase 10.1063/1.5007042} {\bibfield  {journal} {\bibinfo  {journal} {Physics of Plasmas}\ }\textbf {\bibinfo {volume} {25}},\ \bibinfo {pages} {032507} (\bibinfo {year} {2018})}\BibitemShut {NoStop}%
\bibitem [{\citenamefont {Goodall}\ and\ \citenamefont {Wesson}(1984)}]{goodall_cine_1984}%
  \BibitemOpen
  \bibfield  {author} {\bibinfo {author} {\bibnamefont {Goodall}, \bibfnamefont {D.~H.~J.}}\ and\ \bibinfo {author} {\bibnamefont {Wesson}, \bibfnamefont {J.~A.}},\ }\bibfield  {title} {\enquote {\bibinfo {title} {Cine observations of {MHD} instabilities in a {Tokamak}},}\ }\href {\doibase 10.1088/0741-3335/26/6/002} {\bibfield  {journal} {\bibinfo  {journal} {Plasma Physics and Controlled Fusion}\ }\textbf {\bibinfo {volume} {26}},\ \bibinfo {pages} {789--797} (\bibinfo {year} {1984})}\BibitemShut {NoStop}%
\bibitem [{\citenamefont {Hansen}\ \emph {et~al.}(2024)\citenamefont {Hansen}, \citenamefont {Stewart}, \citenamefont {Burgess}, \citenamefont {Pharr}, \citenamefont {Guizzo}, \citenamefont {Logak}, \citenamefont {Nelson},\ and\ \citenamefont {Paz-Soldan}}]{hansen_tokamaker_2024}%
  \BibitemOpen
  \bibfield  {author} {\bibinfo {author} {\bibnamefont {Hansen}, \bibfnamefont {C.}}, \bibinfo {author} {\bibnamefont {Stewart}, \bibfnamefont {I.~G.}}, \bibinfo {author} {\bibnamefont {Burgess}, \bibfnamefont {D.}}, \bibinfo {author} {\bibnamefont {Pharr}, \bibfnamefont {M.}}, \bibinfo {author} {\bibnamefont {Guizzo}, \bibfnamefont {S.}}, \bibinfo {author} {\bibnamefont {Logak}, \bibfnamefont {F.}}, \bibinfo {author} {\bibnamefont {Nelson}, \bibfnamefont {A.~O.}}, \ and\ \bibinfo {author} {\bibnamefont {Paz-Soldan}, \bibfnamefont {C.}},\ }\bibfield  {title} {\enquote {\bibinfo {title} {{TokaMaker}: {An} open-source time-dependent {Grad}-{Shafranov} tool for the design and modeling of axisymmetric fusion devices},}\ }\href {\doibase https://doi.org/10.1016/j.cpc.2024.109111} {\bibfield  {journal} {\bibinfo  {journal} {Computer Physics Communications}\ }\textbf {\bibinfo {volume} {298}},\ \bibinfo {pages} {109111} (\bibinfo {year} {2024})}\BibitemShut {NoStop}%
\bibitem [{\citenamefont {Hindmarsh}(1983)}]{hindmarsh_odepack_1983}%
  \BibitemOpen
  \bibfield  {author} {\bibinfo {author} {\bibnamefont {Hindmarsh}, \bibfnamefont {A.~C.}},\ }\bibfield  {title} {\enquote {\bibinfo {title} {{ODEPACK}, a systematized collection of {ODE} solvers},}\ }\href@noop {} {\bibfield  {journal} {\bibinfo  {journal} {Scientific Computing edited by R. Stepleman}\ } (\bibinfo {year} {1983})}\BibitemShut {NoStop}%
\bibitem [{\citenamefont {King}\ \emph {et~al.}(2015)\citenamefont {King}, \citenamefont {Strait}, \citenamefont {Lazerson}, \citenamefont {Ferraro}, \citenamefont {Logan}, \citenamefont {Haskey}, \citenamefont {Park}, \citenamefont {Hanson}, \citenamefont {Lanctot}, \citenamefont {Liu}, \citenamefont {Nazikian}, \citenamefont {Okabayashi}, \citenamefont {Paz-Soldan}, \citenamefont {Shiraki},\ and\ \citenamefont {Turnbull}}]{king_experimental_2015}%
  \BibitemOpen
  \bibfield  {author} {\bibinfo {author} {\bibnamefont {King}, \bibfnamefont {J.~D.}}, \bibinfo {author} {\bibnamefont {Strait}, \bibfnamefont {E.~J.}}, \bibinfo {author} {\bibnamefont {Lazerson}, \bibfnamefont {S.~A.}}, \bibinfo {author} {\bibnamefont {Ferraro}, \bibfnamefont {N.~M.}}, \bibinfo {author} {\bibnamefont {Logan}, \bibfnamefont {N.~C.}}, \bibinfo {author} {\bibnamefont {Haskey}, \bibfnamefont {S.~R.}}, \bibinfo {author} {\bibnamefont {Park}, \bibfnamefont {J.-K.}}, \bibinfo {author} {\bibnamefont {Hanson}, \bibfnamefont {J.~M.}}, \bibinfo {author} {\bibnamefont {Lanctot}, \bibfnamefont {M.~J.}}, \bibinfo {author} {\bibnamefont {Liu}, \bibfnamefont {Y.}}, \bibinfo {author} {\bibnamefont {Nazikian}, \bibfnamefont {R.}}, \bibinfo {author} {\bibnamefont {Okabayashi}, \bibfnamefont {M.}}, \bibinfo {author} {\bibnamefont {Paz-Soldan}, \bibfnamefont {C.}}, \bibinfo {author} {\bibnamefont {Shiraki}, \bibfnamefont {D.}}, \ and\ \bibinfo {author} {\bibnamefont {Turnbull}, \bibfnamefont {A.~D.}},\
  }\bibfield  {title} {\enquote {\bibinfo {title} {Experimental tests of linear and nonlinear three-dimensional equilibrium models in {DIII}-{D}},}\ }\href {\doibase 10.1063/1.4923017} {\bibfield  {journal} {\bibinfo  {journal} {Physics of Plasmas}\ }\textbf {\bibinfo {volume} {22}},\ \bibinfo {pages} {072501} (\bibinfo {year} {2015})}\BibitemShut {NoStop}%
\bibitem [{\citenamefont {Lao}\ \emph {et~al.}(2005)\citenamefont {Lao}, \citenamefont {John}, \citenamefont {Peng}, \citenamefont {Ferron}, \citenamefont {Strait}, \citenamefont {Taylor}, \citenamefont {Meyer}, \citenamefont {Zhang},\ and\ \citenamefont {You}}]{lao_mhd_2005}%
  \BibitemOpen
  \bibfield  {author} {\bibinfo {author} {\bibnamefont {Lao}, \bibfnamefont {L.~L.}}, \bibinfo {author} {\bibnamefont {John}, \bibfnamefont {H.~E.~S.}}, \bibinfo {author} {\bibnamefont {Peng}, \bibfnamefont {Q.}}, \bibinfo {author} {\bibnamefont {Ferron}, \bibfnamefont {J.~R.}}, \bibinfo {author} {\bibnamefont {Strait}, \bibfnamefont {E.~J.}}, \bibinfo {author} {\bibnamefont {Taylor}, \bibfnamefont {T.~S.}}, \bibinfo {author} {\bibnamefont {Meyer}, \bibfnamefont {W.~H.}}, \bibinfo {author} {\bibnamefont {Zhang}, \bibfnamefont {C.}}, \ and\ \bibinfo {author} {\bibnamefont {You}, \bibfnamefont {K.~I.}},\ }\bibfield  {title} {\enquote {\bibinfo {title} {{MHD} {Equilibrium} {Reconstruction} in the {DIII}-{D} {Tokamak}},}\ }\href {\doibase 10.13182/FST48-968} {\bibfield  {journal} {\bibinfo  {journal} {Fusion Science and Technology}\ }\textbf {\bibinfo {volume} {48}},\ \bibinfo {pages} {968--977} (\bibinfo {year} {2005})}\BibitemShut {NoStop}%
\bibitem [{\citenamefont {Levesque}\ \emph {et~al.}(2013)\citenamefont {Levesque}, \citenamefont {Rath}, \citenamefont {Shiraki}, \citenamefont {Angelini}, \citenamefont {Bialek}, \citenamefont {Byrne}, \citenamefont {DeBono}, \citenamefont {Hughes}, \citenamefont {Mauel}, \citenamefont {Navratil}, \citenamefont {Peng}, \citenamefont {Rhodes},\ and\ \citenamefont {Stoafer}}]{levesque_multimode_2013}%
  \BibitemOpen
  \bibfield  {author} {\bibinfo {author} {\bibnamefont {Levesque}, \bibfnamefont {J.}}, \bibinfo {author} {\bibnamefont {Rath}, \bibfnamefont {N.}}, \bibinfo {author} {\bibnamefont {Shiraki}, \bibfnamefont {D.}}, \bibinfo {author} {\bibnamefont {Angelini}, \bibfnamefont {S.}}, \bibinfo {author} {\bibnamefont {Bialek}, \bibfnamefont {J.}}, \bibinfo {author} {\bibnamefont {Byrne}, \bibfnamefont {P.}}, \bibinfo {author} {\bibnamefont {DeBono}, \bibfnamefont {B.}}, \bibinfo {author} {\bibnamefont {Hughes}, \bibfnamefont {P.}}, \bibinfo {author} {\bibnamefont {Mauel}, \bibfnamefont {M.}}, \bibinfo {author} {\bibnamefont {Navratil}, \bibfnamefont {G.}}, \bibinfo {author} {\bibnamefont {Peng}, \bibfnamefont {Q.}}, \bibinfo {author} {\bibnamefont {Rhodes}, \bibfnamefont {D.}}, \ and\ \bibinfo {author} {\bibnamefont {Stoafer}, \bibfnamefont {C.}},\ }\bibfield  {title} {\enquote {\bibinfo {title} {Multimode observations and {3D} magnetic control of the boundary of a tokamak plasma},}\ }\href {\doibase
  10.1088/0029-5515/53/7/073037} {\bibfield  {journal} {\bibinfo  {journal} {Nuclear Fusion}\ }\textbf {\bibinfo {volume} {53}},\ \bibinfo {pages} {073037} (\bibinfo {year} {2013})}\BibitemShut {NoStop}%
\bibitem [{\citenamefont {Li}\ \emph {et~al.}(2016)\citenamefont {Li}, \citenamefont {Liu}, \citenamefont {Kirk}, \citenamefont {Wang}, \citenamefont {Liang}, \citenamefont {Ryan}, \citenamefont {Suttrop}, \citenamefont {Dunne}, \citenamefont {Fischer}, \citenamefont {Fuchs}, \citenamefont {Kurzan}, \citenamefont {Piovesan}, \citenamefont {Willensdorfer},\ and\ \citenamefont {Zhong}}]{li_modelling_2016}%
  \BibitemOpen
  \bibfield  {author} {\bibinfo {author} {\bibnamefont {Li}, \bibfnamefont {L.}}, \bibinfo {author} {\bibnamefont {Liu}, \bibfnamefont {Y.}}, \bibinfo {author} {\bibnamefont {Kirk}, \bibfnamefont {A.}}, \bibinfo {author} {\bibnamefont {Wang}, \bibfnamefont {N.}}, \bibinfo {author} {\bibnamefont {Liang}, \bibfnamefont {Y.}}, \bibinfo {author} {\bibnamefont {Ryan}, \bibfnamefont {D.}}, \bibinfo {author} {\bibnamefont {Suttrop}, \bibfnamefont {W.}}, \bibinfo {author} {\bibnamefont {Dunne}, \bibfnamefont {M.}}, \bibinfo {author} {\bibnamefont {Fischer}, \bibfnamefont {R.}}, \bibinfo {author} {\bibnamefont {Fuchs}, \bibfnamefont {J.}}, \bibinfo {author} {\bibnamefont {Kurzan}, \bibfnamefont {B.}}, \bibinfo {author} {\bibnamefont {Piovesan}, \bibfnamefont {P.}}, \bibinfo {author} {\bibnamefont {Willensdorfer}, \bibfnamefont {M.}}, \ and\ \bibinfo {author} {\bibnamefont {Zhong}, \bibfnamefont {F.}},\ }\bibfield  {title} {\enquote {\bibinfo {title} {Modelling plasma response to {RMP} fields in {ASDEX} {Upgrade} with
  varying edge safety factor and triangularity},}\ }\href {\doibase 10.1088/0029-5515/56/12/126007} {\bibfield  {journal} {\bibinfo  {journal} {Nuclear Fusion}\ }\textbf {\bibinfo {volume} {56}},\ \bibinfo {pages} {126007} (\bibinfo {year} {2016})}\BibitemShut {NoStop}%
\bibitem [{\citenamefont {Li}\ \emph {et~al.}(2021)\citenamefont {Li}, \citenamefont {Chen}, \citenamefont {Fan}, \citenamefont {Zhu}, \citenamefont {Huang}, \citenamefont {Wen}, \citenamefont {He}, \citenamefont {Yang},\ and\ \citenamefont {Yin}}]{li_development_2021}%
  \BibitemOpen
  \bibfield  {author} {\bibinfo {author} {\bibnamefont {Li}, \bibfnamefont {X.}}, \bibinfo {author} {\bibnamefont {Chen}, \bibfnamefont {C.}}, \bibinfo {author} {\bibnamefont {Fan}, \bibfnamefont {W.}}, \bibinfo {author} {\bibnamefont {Zhu}, \bibfnamefont {R.}}, \bibinfo {author} {\bibnamefont {Huang}, \bibfnamefont {S.}}, \bibinfo {author} {\bibnamefont {Wen}, \bibfnamefont {X.}}, \bibinfo {author} {\bibnamefont {He}, \bibfnamefont {Z.}}, \bibinfo {author} {\bibnamefont {Yang}, \bibfnamefont {Q.}}, \ and\ \bibinfo {author} {\bibnamefont {Yin}, \bibfnamefont {Z.}},\ }\bibfield  {title} {\enquote {\bibinfo {title} {Development of a real-time magnetic island reconstruction system based on {PCIe} platform for {HL}-{2A} tokamak},}\ }\href {\doibase 10.1088/2058-6272/ac0ab7} {\bibfield  {journal} {\bibinfo  {journal} {Plasma Science and Technology}\ }\textbf {\bibinfo {volume} {23}},\ \bibinfo {pages} {085103} (\bibinfo {year} {2021})}\BibitemShut {NoStop}%
\bibitem [{\citenamefont {Liu}\ \emph {et~al.}(2000)\citenamefont {Liu}, \citenamefont {Bondeson}, \citenamefont {Fransson}, \citenamefont {Lennartson},\ and\ \citenamefont {Breitholtz}}]{liu_feedback_2000}%
  \BibitemOpen
  \bibfield  {author} {\bibinfo {author} {\bibnamefont {Liu}, \bibfnamefont {Y.~Q.}}, \bibinfo {author} {\bibnamefont {Bondeson}, \bibfnamefont {A.}}, \bibinfo {author} {\bibnamefont {Fransson}, \bibfnamefont {C.~M.}}, \bibinfo {author} {\bibnamefont {Lennartson}, \bibfnamefont {B.}}, \ and\ \bibinfo {author} {\bibnamefont {Breitholtz}, \bibfnamefont {C.}},\ }\bibfield  {title} {\enquote {\bibinfo {title} {Feedback stabilization of nonaxisymmetric resistive wall modes in tokamaks. {I}. {Electromagnetic} model},}\ }\href {\doibase 10.1063/1.1287744} {\bibfield  {journal} {\bibinfo  {journal} {Physics of Plasmas}\ }\textbf {\bibinfo {volume} {7}},\ \bibinfo {pages} {3681--3690} (\bibinfo {year} {2000})}\BibitemShut {NoStop}%
\bibitem [{\citenamefont {Logan}\ \emph {et~al.}(2016)\citenamefont {Logan}, \citenamefont {Park}, \citenamefont {Paz-Soldan}, \citenamefont {Lanctot}, \citenamefont {Smith},\ and\ \citenamefont {Burrell}}]{logan_dependence_2016}%
  \BibitemOpen
  \bibfield  {author} {\bibinfo {author} {\bibnamefont {Logan}, \bibfnamefont {N.}}, \bibinfo {author} {\bibnamefont {Park}, \bibfnamefont {J.-K.}}, \bibinfo {author} {\bibnamefont {Paz-Soldan}, \bibfnamefont {C.}}, \bibinfo {author} {\bibnamefont {Lanctot}, \bibfnamefont {M.}}, \bibinfo {author} {\bibnamefont {Smith}, \bibfnamefont {S.}}, \ and\ \bibinfo {author} {\bibnamefont {Burrell}, \bibfnamefont {K.}},\ }\bibfield  {title} {\enquote {\bibinfo {title} {Dependence of neoclassical toroidal viscosity on the poloidal spectrum of applied nonaxisymmetric fields},}\ }\href {\doibase 10.1088/0029-5515/56/3/036008} {\bibfield  {journal} {\bibinfo  {journal} {Nuclear Fusion}\ }\textbf {\bibinfo {volume} {56}},\ \bibinfo {pages} {036008} (\bibinfo {year} {2016})}\BibitemShut {NoStop}%
\bibitem [{\citenamefont {Logan}\ \emph {et~al.}(2013)\citenamefont {Logan}, \citenamefont {Park}, \citenamefont {Kim}, \citenamefont {Wang},\ and\ \citenamefont {Berkery}}]{logan_neoclassical_2013}%
  \BibitemOpen
  \bibfield  {author} {\bibinfo {author} {\bibnamefont {Logan}, \bibfnamefont {N.~C.}}, \bibinfo {author} {\bibnamefont {Park}, \bibfnamefont {J.-K.}}, \bibinfo {author} {\bibnamefont {Kim}, \bibfnamefont {K.}}, \bibinfo {author} {\bibnamefont {Wang}, \bibfnamefont {Z.}}, \ and\ \bibinfo {author} {\bibnamefont {Berkery}, \bibfnamefont {J.~W.}},\ }\bibfield  {title} {\enquote {\bibinfo {title} {Neoclassical toroidal viscosity in perturbed equilibria with general tokamak geometry},}\ }\href {\doibase 10.1063/1.4849395} {\bibfield  {journal} {\bibinfo  {journal} {Physics of Plasmas}\ }\textbf {\bibinfo {volume} {20}},\ \bibinfo {pages} {122507} (\bibinfo {year} {2013})}\BibitemShut {NoStop}%
\bibitem [{\citenamefont {Makishima}\ \emph {et~al.}(1976)\citenamefont {Makishima}, \citenamefont {Tominaga}, \citenamefont {Tohyama},\ and\ \citenamefont {Yoshikawa}}]{makishima_simultaneous_1976}%
  \BibitemOpen
  \bibfield  {author} {\bibinfo {author} {\bibnamefont {Makishima}, \bibfnamefont {K.}}, \bibinfo {author} {\bibnamefont {Tominaga}, \bibfnamefont {T.}}, \bibinfo {author} {\bibnamefont {Tohyama}, \bibfnamefont {H.}}, \ and\ \bibinfo {author} {\bibnamefont {Yoshikawa}, \bibfnamefont {S.}},\ }\bibfield  {title} {\enquote {\bibinfo {title} {Simultaneous {Measurements} of the {Plasma} {Current} {Profile} and {Instabilities} in a {Tokamak}},}\ }\href {\doibase 10.1103/PhysRevLett.36.142} {\bibfield  {journal} {\bibinfo  {journal} {Physical Review Letters}\ }\textbf {\bibinfo {volume} {36}},\ \bibinfo {pages} {142--145} (\bibinfo {year} {1976})}\BibitemShut {NoStop}%
\bibitem [{\citenamefont {Mirnov}\ and\ \citenamefont {Semenov}(1971)}]{mirnov_investigation_1971}%
  \BibitemOpen
  \bibfield  {author} {\bibinfo {author} {\bibnamefont {Mirnov}, \bibfnamefont {S.~V.}}\ and\ \bibinfo {author} {\bibnamefont {Semenov}, \bibfnamefont {I.~B.}},\ }\bibfield  {title} {\enquote {\bibinfo {title} {Investigation of the instabilities of the plasma string in the {Tokamak}-3 system by means of a correlation method},}\ }\href {\doibase 10.1007/BF01788387} {\bibfield  {journal} {\bibinfo  {journal} {Soviet Atomic Energy}\ }\textbf {\bibinfo {volume} {30}},\ \bibinfo {pages} {22--29} (\bibinfo {year} {1971})}\BibitemShut {NoStop}%
\bibitem [{\citenamefont {Park}, \citenamefont {Boozer},\ and\ \citenamefont {Glasser}(2007)}]{park_computation_2007}%
  \BibitemOpen
  \bibfield  {author} {\bibinfo {author} {\bibnamefont {Park}, \bibfnamefont {J.-k.}}, \bibinfo {author} {\bibnamefont {Boozer}, \bibfnamefont {A.~H.}}, \ and\ \bibinfo {author} {\bibnamefont {Glasser}, \bibfnamefont {A.~H.}},\ }\bibfield  {title} {\enquote {\bibinfo {title} {Computation of three-dimensional tokamak and spherical torus equilibria},}\ }\href {\doibase 10.1063/1.2732170} {\bibfield  {journal} {\bibinfo  {journal} {Physics of Plasmas}\ }\textbf {\bibinfo {volume} {14}},\ \bibinfo {pages} {052110} (\bibinfo {year} {2007})}\BibitemShut {NoStop}%
\bibitem [{\citenamefont {Park}, \citenamefont {Boozer},\ and\ \citenamefont {Menard}(2008)}]{park_spectral_2008}%
  \BibitemOpen
  \bibfield  {author} {\bibinfo {author} {\bibnamefont {Park}, \bibfnamefont {J.-k.}}, \bibinfo {author} {\bibnamefont {Boozer}, \bibfnamefont {A.~H.}}, \ and\ \bibinfo {author} {\bibnamefont {Menard}, \bibfnamefont {J.~E.}},\ }\bibfield  {title} {\enquote {\bibinfo {title} {Spectral asymmetry due to magnetic coordinates},}\ }\href {\doibase 10.1063/1.2932110} {\bibfield  {journal} {\bibinfo  {journal} {Physics of Plasmas}\ }\textbf {\bibinfo {volume} {15}},\ \bibinfo {pages} {064501} (\bibinfo {year} {2008})}\BibitemShut {NoStop}%
\bibitem [{\citenamefont {Park}\ \emph {et~al.}(2009)\citenamefont {Park}, \citenamefont {Boozer}, \citenamefont {Menard}, \citenamefont {Garofalo}, \citenamefont {Schaffer}, \citenamefont {Hawryluk}, \citenamefont {Kaye}, \citenamefont {Gerhardt}, \citenamefont {Sabbagh},\ and\ \citenamefont {{NSTX Team}}}]{park_importance_2009}%
  \BibitemOpen
  \bibfield  {author} {\bibinfo {author} {\bibnamefont {Park}, \bibfnamefont {J.-k.}}, \bibinfo {author} {\bibnamefont {Boozer}, \bibfnamefont {A.~H.}}, \bibinfo {author} {\bibnamefont {Menard}, \bibfnamefont {J.~E.}}, \bibinfo {author} {\bibnamefont {Garofalo}, \bibfnamefont {A.~M.}}, \bibinfo {author} {\bibnamefont {Schaffer}, \bibfnamefont {M.~J.}}, \bibinfo {author} {\bibnamefont {Hawryluk}, \bibfnamefont {R.~J.}}, \bibinfo {author} {\bibnamefont {Kaye}, \bibfnamefont {S.~M.}}, \bibinfo {author} {\bibnamefont {Gerhardt}, \bibfnamefont {S.~P.}}, \bibinfo {author} {\bibnamefont {Sabbagh}, \bibfnamefont {S.~A.}}, \ and\ \bibinfo {author} {\bibnamefont {{NSTX Team}},},\ }\bibfield  {title} {\enquote {\bibinfo {title} {Importance of plasma response to nonaxisymmetric perturbations in tokamaks},}\ }\href {\doibase 10.1063/1.3122862} {\bibfield  {journal} {\bibinfo  {journal} {Physics of Plasmas}\ }\textbf {\bibinfo {volume} {16}},\ \bibinfo {pages} {056115} (\bibinfo {year} {2009})}\BibitemShut {NoStop}%
\bibitem [{\citenamefont {Park}\ and\ \citenamefont {Logan}(2017)}]{park_self-consistent_2017}%
  \BibitemOpen
  \bibfield  {author} {\bibinfo {author} {\bibnamefont {Park}, \bibfnamefont {J.-K.}}\ and\ \bibinfo {author} {\bibnamefont {Logan}, \bibfnamefont {N.~C.}},\ }\bibfield  {title} {\enquote {\bibinfo {title} {Self-consistent perturbed equilibrium with neoclassical toroidal torque in tokamaks},}\ }\href {\doibase 10.1063/1.4977898} {\bibfield  {journal} {\bibinfo  {journal} {Physics of Plasmas}\ }\textbf {\bibinfo {volume} {24}},\ \bibinfo {pages} {032505} (\bibinfo {year} {2017})}\BibitemShut {NoStop}%
\bibitem [{\citenamefont {Paz-Soldan}\ \emph {et~al.}(2016)\citenamefont {Paz-Soldan}, \citenamefont {Logan}, \citenamefont {Haskey}, \citenamefont {Nazikian}, \citenamefont {Strait}, \citenamefont {Chen}, \citenamefont {Ferraro}, \citenamefont {King}, \citenamefont {Lyons},\ and\ \citenamefont {Park}}]{paz-soldan_equilibrium_2016}%
  \BibitemOpen
  \bibfield  {author} {\bibinfo {author} {\bibnamefont {Paz-Soldan}, \bibfnamefont {C.}}, \bibinfo {author} {\bibnamefont {Logan}, \bibfnamefont {N.}}, \bibinfo {author} {\bibnamefont {Haskey}, \bibfnamefont {S.}}, \bibinfo {author} {\bibnamefont {Nazikian}, \bibfnamefont {R.}}, \bibinfo {author} {\bibnamefont {Strait}, \bibfnamefont {E.}}, \bibinfo {author} {\bibnamefont {Chen}, \bibfnamefont {X.}}, \bibinfo {author} {\bibnamefont {Ferraro}, \bibfnamefont {N.}}, \bibinfo {author} {\bibnamefont {King}, \bibfnamefont {J.}}, \bibinfo {author} {\bibnamefont {Lyons}, \bibfnamefont {B.}}, \ and\ \bibinfo {author} {\bibnamefont {Park}, \bibfnamefont {J.-K.}},\ }\bibfield  {title} {\enquote {\bibinfo {title} {Equilibrium drives of the low and high field side n = 2 plasma response and impact on global confinement},}\ }\href {\doibase 10.1088/0029-5515/56/5/056001} {\bibfield  {journal} {\bibinfo  {journal} {Nuclear Fusion}\ }\textbf {\bibinfo {volume} {56}},\ \bibinfo {pages} {056001} (\bibinfo {year}
  {2016})}\BibitemShut {NoStop}%
\bibitem [{\citenamefont {Pletzer}, \citenamefont {Bondeson},\ and\ \citenamefont {Dewar}(1994)}]{pletzer_linear_1994}%
  \BibitemOpen
  \bibfield  {author} {\bibinfo {author} {\bibnamefont {Pletzer}, \bibfnamefont {A.}}, \bibinfo {author} {\bibnamefont {Bondeson}, \bibfnamefont {A.}}, \ and\ \bibinfo {author} {\bibnamefont {Dewar}, \bibfnamefont {R.~L.}},\ }\bibfield  {title} {\enquote {\bibinfo {title} {Linear {Stability} of {Resistive} {MHD} {Modes}: {Axisymmetric} {Toroidal} {Computation} of the {Outer} {Region} {Matching} {Data}},}\ }\href {\doibase https://doi.org/10.1006/jcph.1994.1215} {\bibfield  {journal} {\bibinfo  {journal} {Journal of Computational Physics}\ }\textbf {\bibinfo {volume} {115}},\ \bibinfo {pages} {530--549} (\bibinfo {year} {1994})}\BibitemShut {NoStop}%
\bibitem [{\citenamefont {Turnbull}\ \emph {et~al.}(2016)\citenamefont {Turnbull}, \citenamefont {Hanson}, \citenamefont {Turco}, \citenamefont {Ferraro}, \citenamefont {Lanctot}, \citenamefont {Lao}, \citenamefont {Strait}, \citenamefont {Piovesan},\ and\ \citenamefont {Martin}}]{turnbull_external_2016}%
  \BibitemOpen
  \bibfield  {author} {\bibinfo {author} {\bibnamefont {Turnbull}, \bibfnamefont {A.~D.}}, \bibinfo {author} {\bibnamefont {Hanson}, \bibfnamefont {J.~M.}}, \bibinfo {author} {\bibnamefont {Turco}, \bibfnamefont {F.}}, \bibinfo {author} {\bibnamefont {Ferraro}, \bibfnamefont {N.~M.}}, \bibinfo {author} {\bibnamefont {Lanctot}, \bibfnamefont {M.~J.}}, \bibinfo {author} {\bibnamefont {Lao}, \bibfnamefont {L.~L.}}, \bibinfo {author} {\bibnamefont {Strait}, \bibfnamefont {E.~J.}}, \bibinfo {author} {\bibnamefont {Piovesan}, \bibfnamefont {P.}}, \ and\ \bibinfo {author} {\bibnamefont {Martin}, \bibfnamefont {P.}},\ }\bibfield  {title} {\enquote {\bibinfo {title} {The external kink mode in diverted tokamaks},}\ }\href {\doibase 10.1017/S0022377816000568} {\bibfield  {journal} {\bibinfo  {journal} {Journal of Plasma Physics}\ }\textbf {\bibinfo {volume} {82}},\ \bibinfo {pages} {515820301} (\bibinfo {year} {2016})}\BibitemShut {NoStop}%
\bibitem [{\citenamefont {Wesson}(1978)}]{wesson_hydromagnetic_1978}%
  \BibitemOpen
  \bibfield  {author} {\bibinfo {author} {\bibnamefont {Wesson}, \bibfnamefont {J.}},\ }\bibfield  {title} {\enquote {\bibinfo {title} {Hydromagnetic stability of tokamaks},}\ }\href {\doibase 10.1088/0029-5515/18/1/010} {\bibfield  {journal} {\bibinfo  {journal} {Nuclear Fusion}\ }\textbf {\bibinfo {volume} {18}},\ \bibinfo {pages} {87--132} (\bibinfo {year} {1978})}\BibitemShut {NoStop}%
\bibitem [{\citenamefont {Xie}\ \emph {et~al.}(2023)\citenamefont {Xie}, \citenamefont {Sun}, \citenamefont {Ma}, \citenamefont {Gu}, \citenamefont {Liu}, \citenamefont {Jia}, \citenamefont {Loarte}, \citenamefont {Wu}, \citenamefont {Chang}, \citenamefont {Jia}, \citenamefont {Zhang}, \citenamefont {Zhou}, \citenamefont {Zang}, \citenamefont {Lyu}, \citenamefont {Fu}, \citenamefont {Sheng}, \citenamefont {Ye}, \citenamefont {Yang}, \citenamefont {Wang},\ and\ \citenamefont {{the EAST Team}}}]{xie_extension_2023}%
  \BibitemOpen
  \bibfield  {author} {\bibinfo {author} {\bibnamefont {Xie}, \bibfnamefont {P.}}, \bibinfo {author} {\bibnamefont {Sun}, \bibfnamefont {Y.}}, \bibinfo {author} {\bibnamefont {Ma}, \bibfnamefont {Q.}}, \bibinfo {author} {\bibnamefont {Gu}, \bibfnamefont {S.}}, \bibinfo {author} {\bibnamefont {Liu}, \bibfnamefont {Y.}}, \bibinfo {author} {\bibnamefont {Jia}, \bibfnamefont {M.}}, \bibinfo {author} {\bibnamefont {Loarte}, \bibfnamefont {A.}}, \bibinfo {author} {\bibnamefont {Wu}, \bibfnamefont {X.}}, \bibinfo {author} {\bibnamefont {Chang}, \bibfnamefont {Y.}}, \bibinfo {author} {\bibnamefont {Jia}, \bibfnamefont {T.}}, \bibinfo {author} {\bibnamefont {Zhang}, \bibfnamefont {T.}}, \bibinfo {author} {\bibnamefont {Zhou}, \bibfnamefont {Z.}}, \bibinfo {author} {\bibnamefont {Zang}, \bibfnamefont {Q.}}, \bibinfo {author} {\bibnamefont {Lyu}, \bibfnamefont {B.}}, \bibinfo {author} {\bibnamefont {Fu}, \bibfnamefont {S.}}, \bibinfo {author} {\bibnamefont {Sheng}, \bibfnamefont {H.}}, \bibinfo {author} {\bibnamefont
  {Ye}, \bibfnamefont {C.}}, \bibinfo {author} {\bibnamefont {Yang}, \bibfnamefont {H.}}, \bibinfo {author} {\bibnamefont {Wang}, \bibfnamefont {H.}}, \ and\ \bibinfo {author} {\bibnamefont {{the EAST Team}},},\ }\bibfield  {title} {\enquote {\bibinfo {title} {Extension of {ELM} suppression window using n = 4 {RMPs} in {EAST}},}\ }\href {\doibase 10.1088/1741-4326/aceb07} {\bibfield  {journal} {\bibinfo  {journal} {Nuclear Fusion}\ }\textbf {\bibinfo {volume} {63}},\ \bibinfo {pages} {096025} (\bibinfo {year} {2023})}\BibitemShut {NoStop}%
\bibitem [{\citenamefont {Yang}\ \emph {et~al.}(2019)\citenamefont {Yang}, \citenamefont {Liu}, \citenamefont {Paz-Soldan}, \citenamefont {Zhou}, \citenamefont {Li}, \citenamefont {Xia}, \citenamefont {He},\ and\ \citenamefont {Wang}}]{yang_resistive_2019}%
  \BibitemOpen
  \bibfield  {author} {\bibinfo {author} {\bibnamefont {Yang}, \bibfnamefont {X.}}, \bibinfo {author} {\bibnamefont {Liu}, \bibfnamefont {Y.}}, \bibinfo {author} {\bibnamefont {Paz-Soldan}, \bibfnamefont {C.}}, \bibinfo {author} {\bibnamefont {Zhou}, \bibfnamefont {L.}}, \bibinfo {author} {\bibnamefont {Li}, \bibfnamefont {L.}}, \bibinfo {author} {\bibnamefont {Xia}, \bibfnamefont {G.}}, \bibinfo {author} {\bibnamefont {He}, \bibfnamefont {Y.}}, \ and\ \bibinfo {author} {\bibnamefont {Wang}, \bibfnamefont {S.}},\ }\bibfield  {title} {\enquote {\bibinfo {title} {Resistive versus ideal plasma response to {RMP} fields in {DIII}-{D}: roles of \textit{q}$_{\textrm{95}}$ and {X}-point geometry},}\ }\href {\doibase 10.1088/1741-4326/ab20f9} {\bibfield  {journal} {\bibinfo  {journal} {Nuclear Fusion}\ }\textbf {\bibinfo {volume} {59}},\ \bibinfo {pages} {086012} (\bibinfo {year} {2019})}\BibitemShut {NoStop}%
\bibitem [{\citenamefont {Zheng}\ \emph {et~al.}(2025)\citenamefont {Zheng}, \citenamefont {Kotschenreuther}, \citenamefont {Waelbroeck},\ and\ \citenamefont {Austin}}]{zheng_x-point_2025}%
  \BibitemOpen
  \bibfield  {author} {\bibinfo {author} {\bibnamefont {Zheng}, \bibfnamefont {L.}}, \bibinfo {author} {\bibnamefont {Kotschenreuther}, \bibfnamefont {M.~T.}}, \bibinfo {author} {\bibnamefont {Waelbroeck}, \bibfnamefont {F.~L.}}, \ and\ \bibinfo {author} {\bibnamefont {Austin}, \bibfnamefont {M.~E.}},\ }\bibfield  {title} {\enquote {\bibinfo {title} {X-point effects on the ideal {MHD} modes in tokamaks in the description of dual-poloidal-region safety factor},}\ }\href {\doibase 10.1063/5.0227541} {\bibfield  {journal} {\bibinfo  {journal} {Physics of Plasmas}\ }\textbf {\bibinfo {volume} {32}},\ \bibinfo {pages} {012501} (\bibinfo {year} {2025})}\BibitemShut {NoStop}%
\end{thebibliography}%
\end{document}